\documentclass[prd,twocolumn,superscriptaddress,floatfix,amsmath,amssymb,amsfonts,nofootinbib,longbibliography]{revtex4-2}

\usepackage{float} 
\usepackage{scalerel}
\usepackage[normalem]{ulem}
\usepackage[english]{babel}
\usepackage{graphicx}
\usepackage{dcolumn}
\usepackage{bm}
\usepackage{blindtext}
\usepackage{verbatim}
\usepackage{relsize}
\usepackage{mathrsfs}
\usepackage{musicography}
\usepackage{amsmath}
\usepackage{blindtext}
\usepackage{cancel}
\usepackage{physics}
\usepackage{epstopdf}
\usepackage{mathtools}
\usepackage{blindtext}
\usepackage{tensor}
\usepackage{color}
\usepackage[usenames,dvipsnames]{pstricks}
\usepackage{epsfig}
\usepackage{pst-grad}
\usepackage{pst-plot} 
\usepackage{hyperref}
\usepackage{verbatim}
\usepackage{slashed}
\usepackage{dsfont}
\usepackage[english]{babel}
\usepackage{amsmath,amssymb,amsthm} 
\usepackage{enumitem}
\usepackage{lipsum}
\usepackage{makerobust}

\usepackage[caption=false]{subfig}

\usepackage{tikz}
\MakeRobustCommand\rotatebox
\MakeRobustCommand\raisebox
\MakeRobustCommand\sbox
\MakeRobustCommand\resizebox

\newcommand{\w}{\omega}

\newcommand{\bdiamond}{%
   \resizebox{!}{0.7ex}{%
      \rotatebox[origin=c]{45}{$\blacksquare$}%
}}

\newcommand{\bd}{\bdiamond}
\newcommand{\memf}{F^{\bd}}
\newcommand{\memg}{G^{\bd}}

\newcommand{\mf}{\mathsf}

\newcommand{\ii}{\mathrm{i}}

\newcommand{\tc}[1]{\textsc{#1}}

\newcommand{\trr}[1]{\textcolor{red}{#1}}

\newcommand{\sect}[1]{%
\noindent\textbf{\textit{#1---}}%
}

\newlist{arrowlist}{itemize}{1}
\setlist[arrowlist]{label=$\rightarrow$}

\allowdisplaybreaks[1] 

\begin{document}


\title{Local Vacuum Entanglement through Most Entangled Modes}

%

\author{T. Rick Perche}
\email{rick.perche@su.se}

\affiliation{Department of Physics, Stockholm University, SE-106 91 Stockholm, Sweden}
\affiliation{Nordita,
KTH Royal Institute of Technology and Stockholm University,
Hannes Alfvéns väg 12, 23, SE-106 91 Stockholm, Sweden}

\author{Patricia Ribes-Metidieri}
\email{patricia.ribesmetidieri@york.ac.uk}

\affiliation{Department of Mathematics, University of York, Heslington, York YO10 5DD, UK}

\begin{abstract}

    We find the shape of the modes containing most of the entanglement between two disjoint spherical regions in the vacuum of a free massless real scalar field in 1+3 Minkowski spacetime. Using the Klco-Beck-Savage method we approximate the field and momentum profile of the vacuum most entangled modes (MEMs) between two spheres. We define a non-perturbative mode-swap entanglement harvesting protocol where two spacelike separated probes extract the maximal amount of entanglement between two regions. Our results are a first step in the direction of understanding the local entanglement structure of quantum fields, pinpointing the exact degrees of freedom that encode accessible quantum correlations in QFT. 

    
    

    
    
    

   \end{abstract}

\maketitle

\sect{Introduction}The vacuum of quantum field theories (QFTs) is known to maximally violate Bell inequalities between complementary regions~\cite{vacuumBell,summers_bells_1987,summers_maximal_1987,vacuumEntanglement} and to contain entanglement between any two spacelike-separated bounded regions~\cite{RainerEnt,Hollands:2017dov}. However, much of the vacuum entanglement is not operationally accessible: extracting the formally infinite entanglement shared by a region and its complement~\cite{sorkinAreaLawOG,SorkinArea1986,areaLaw1993,kelly,witten,EntEmbezz} would require access to arbitrarily small length scales and, consequently, unbounded energies. Physical probes, by contrast, are restricted to finite-energy couplings and finite, non-complementary regions~\cite{bianchi_entropy_2019, ABRM_ubiquitous,EduAndIvan}. 

This raises a central question: \textit{which local field degrees of freedom encode the entanglement between finite regions?} Despite recent progress, this question remains mostly unanswered. One the one hand, protocols such as entanglement harvesting~\cite{Valentini1991,Reznik2003,reznik2,Nick,Pozas-Kerstjens:2015} demonstrate that localized probes can extract vacuum entanglement, promoting it from a mere academic concept to a physical resource~\cite{Martin-Martinez_Aasen_Kempf_2013,Martin-Martinez_Sutherland_2014,Layden_Martin-Martinez_Kempf_2016,AsplingLawler2024,phil}. On the other hand, with few exceptions~\cite{BeyondPertUDW,Farming,Polo-Gomez:2023gaz}, current theoretical approaches to vacuum entanglement extraction are typically perturbative (see e.g.~\cite{Valentini1991,Reznik2003,reznik2,Nick,Pozas-Kerstjens:2015,NickEdu2014,Pozas2016,HarvestingBHLaura,foo,HarvestingAccelerationRobb,EricksonZero,ericksonNew,carol,boris,quantClass,FullyRelativisticEH}), limiting their ability to \textit{quantify} vacuum entanglement, and present a major challenge in the description of recent experimental proposals~\cite{Mo_Onoe2022RapidlySwitchedUDW,Mo_Settembrini2022VacuumCorrelationsNatCommun,adamExperimentalEH2025,BECEH}. Identifying the field degrees of freedom that encode the entanglement between two regions is the first step to fully understand the local structure of vacuum entanglement and exploit it as a resource.

While only few results on local entanglement in relativistic QFTs are available, valuable intuition can be obtained from simulating massive QFTs as a lattice of interacting harmonic oscillators. In this context, not only can Gaussian techniques be used to estimate the vacuum entanglement between two regions~\cite{gaussEnt,reznikLattice}, but the recently developed Klco-Beck-Savage (KBS) algorithm~\cite{klco_entanglement_2022} is able to identify the individual modes contributing to this entanglement. Importantly, these methods can be applied to a bipartition of any multiparite Gaussian quantum system~\cite{partnerformula}.

In this Letter, we use the KBS method~\cite{klco_entanglement_2022} to obtain the most entangled modes, as quantified by the logarithmic negativity, of the vacuum of a massless scalar field between two spherical regions. These most entangled modes (MEMs) encode the dominant distillable entanglement between the subregions. We then propose an idealized protocol where two spacelike-separated probes couple directly to these modes and efficiently extract their entanglement. Our results both contribute to our understanding of the entanglement structure of quantum fields and pinpoint the exact degrees of freedom that encode accessible quantum correlations in QFT.

\sect{Local degrees of freedom of a QFT}The modern formulation of QFT associates local algebras to each region of spacetime, with each algebra encoding the degrees of freedom associated to the corresponding region. For instance, the QFT for a Klein-Gordon field is formulated by a collection of local algebras, $\mathcal{A}(\mathcal{O})$ associated to each causally complete spacetime region $\mathcal{O}$. Intuitively, each subalgebra $\mathcal{A}(\mathcal{O})$ encodes a representation of the local degrees of freedom of the field supported in the region. The local algebras $\mathcal{A}(\mathcal{O})$ are generated by smeared field observables of the form
\begin{equation}
    \hat \phi(f) = \int \dd V f(\mf x) \hat \phi (\mf x), \text{ with } f\in C_c^\infty(\mathcal{O}),
\end{equation} 
where $\dd V$ is the spacetime volume element. Although the object $\hat \phi (\mf x)$,  heuristically understood as the field operator at a spacetime point, does not define an operator in any reasonable way, it can be formally understood as the kernel of an operator-valued distribution $\hat{\phi}:f\mapsto \hat{\phi}(f)$ (see, e.g.~\cite{Haag,Wald2}).

To explicitly define local modes in QFT, it is often more convenient to work with an equivalent formulation of the theory in the canonical picture, where the field amplitude and momentum at a surface are directly related to initial data for classical solutions. In this construction one considers a one-parameter family of Cauchy hypersurfaces $\Sigma_t$ and, for a given $t$, each subregion $\Sigma_U\subset \Sigma_t$ is associated to a local algebra $\mathcal{A}(U)$, generated by smeared field and momentum operators at the surface. More precisely, the algebra $\mathcal{A}(\Sigma_U)$ is generated by the symbols $\hat{\Phi}(F)$ and $\hat{\Pi}(G)$ acting linearly on real spatial functions $F$ and $G$ supported in $\Sigma_U$ and satisfying the commutation relations:
\begin{equation}
    \begin{aligned}
    [\hat{\Phi}(F&),\hat{\Phi}(G)] = [\hat{\Pi}(F),\hat{\Pi}(G)] = 0, \\
    &[\hat{\Phi}(F),\hat{\Pi}(G)] = i \langle F,G\rangle,
    \end{aligned}
\end{equation}
where $\langle\,\cdot\,,\,\cdot\,\rangle$ denotes the real $L^2$ inner product in $\Sigma_t$. For convenience, we will restrict our attention to the case where $t$ is an inertial time coordinate in Minkowski, and $\Sigma$ is the associated Cauchy surface at $t=0$. In this case, the kernels $\hat{\Phi}(\vec{x})$ and $\hat{\Pi}(\vec{x})$ can then be formally identified with the field amplitude and momentum $\hat{\Phi}(\vec x) := \hat \phi(\mf x)|_{\Sigma}$, $\hat \Pi(\vec x) := \partial_t \hat \phi(\mf x)|_{\Sigma}$, satisfying the standard equal-time commutation relations.


In the canonical formulation, pairs of functions $\bm \gamma := (G(\vec x), F(\vec x ))$ can be identified with initial data for the Klein-Gordon equation,  and hence with elements of the classical phase space $\bm \gamma \in \Gamma$ of the theory. Following~\cite{ABRM_deSitter_long,partnersqft},  it is convenient to identify each linear smeared operator with an element of (the enlargement by Cauchy completion of) $\Gamma$ through 
\begin{equation}
    \hat O_{\gamma} := \bm \w(\bm \gamma, \bm{\hat R}) = \hat \Phi(F) - \hat \Pi(G), 
\end{equation}
where $\bm{\hat R} := (\hat \Phi(\vec x), \hat \Pi(\vec x))$ are formal field and momentum operators in $\Sigma_t$, and $\bm \omega$ denotes the symplectic product in $\Gamma$,\footnote{The vector space structure of the free field theory allows us to identify $\Gamma$ with its tangent space and the symplectic product as acting on elements in $\Gamma$. }  given by 
\begin{equation}
    \bm \omega(\bm\gamma_1,\bm\gamma_2) = \int_{\Sigma}\dd^3x (F_1(\vec x) G_2(\vec x) -G_1(\vec x) F_2(\vec x)),
\end{equation}
where $\bm \gamma_i := (G_i(\vec x),F_i (\vec x))$, for $i=1,2$.

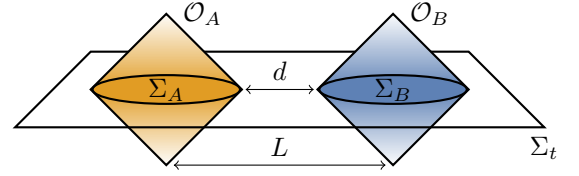
\begin{figure}[h!]
    \centering
\usetikzlibrary{positioning,fadings,through}
\definecolor{wolfram1}{rgb}{0.3686274509803922,0.5058823529411764,0.7098039215686275}
\definecolor{wolfram2}{rgb}{0.8823529411764706,0.611764705882353,0.1411764705882353}

\begin{tikzpicture}
    \node (c) at (0,0) {};

     \draw[thick,wolfram2, fill=wolfram2,path fading=north] (-5/2+0.025, 0) -- (-3/2,1) -- (-1/2-0.025,0) --cycle;
    \draw[thick,wolfram2,fill=wolfram2, path fading=south] (-5/2+0.025, 0) --  (-1/2-0.025,0) -- (-3/2,-1) --cycle;
     \draw[thick,wolfram1, fill=wolfram1,path fading=north] (5/2-0.025, 0) -- (3/2,1) -- (1/2+0.025,0) --cycle;
    \draw[thick,wolfram1,fill=wolfram1, path fading=south] (5/2-0.025, 0) --  (1/2+0.025,0) -- (3/2,-1) --cycle;
        
    \draw[thick] (-2,1/2)--(-5/2,1/2) -- (-7/2, -1/2) -- (7/2,-1/2) -- (5/2,1/2) -- (2,1/2); 
    \draw[thick] (-1,1/2) -- (1,1/2);
    \draw[thick,fill=wolfram2] (-3/2,0) ellipse (0.975 and 0.19); 
    \draw[thick,fill=wolfram1] (3/2,0) ellipse (0.975 and 0.19); 
  
    \draw[thick] (-5/2, 0) -- (-3/2,1) -- (-1/2,0) -- (-3/2,-1) --cycle;
    \draw[thick] (5/2, 0) -- (3/2,1) -- (1/2,0) -- (3/2,-1) --cycle;
    \draw[<->] (-3/2+0.1,-1)--(3/2-0.1,-1); 
    \node[anchor=south] at (0,-1) {$L$};
    \draw[<->]  (-1/2+0.05,0)--(1/2-0.05,0); 
    \node[anchor=south] at (0,0) {$d$}; 
\node[anchor=north] at (7/2,-1/2) {$\Sigma_{t}$}; 
\node at (-3/2,0) {$\Sigma_A$}; 
\node at (3/2,0) {$\Sigma_B$}; 
\node[anchor=west] at (-3/2+1/10,1) {$\mathcal{O}_A$}; 
\node[anchor=west] at (3/2+1/10,1) {$\mathcal{O}_B$}; 

\end{tikzpicture}
    \caption{ Two spherical subregions $\Sigma_\tc{a}$ and $\Sigma_\tc{b}$ at the Cauchy surface $\Sigma_t$ (depicted in 2+1 dimensions), and the associated causal diamonds $\mathcal{O}_\tc{a}$ and $\mathcal{O}_\tc{b}$, where the corresponding covariant algebras of observables are defined.}
    \label{fig:sigmaO}
\end{figure}

The equivalence between the covariantly smeared algebras and the canonical picture is established by the causal propagator $E$ (the unique antisymmetric bi-solution to the equations of motion) through the fact that $\hat{\phi}(f) = - \hat{\Phi}(F) + \hat{\Pi}(G)$, where $G = Ef|_{\Sigma_t}$ and $F = \partial_t Ef|_{\Sigma_t}$, yielding an explicit isomorphism between the algebras $\mathcal{A}(\Sigma_U)$ and $\mathcal{A}(\mathcal{O}_U)$, where $\mathcal{O}_U$ is the causal diamond defined by the spatial subset $\Sigma_U$ (see Fig.~\ref{fig:sigmaO}). 

\newcommand{\patricianocry}{%
\bm \sigma(\bm \gamma,\bm \gamma')%
}

In this context, a ``single-degree of freedom'' or \textit{mode} of the field is identified with the subalgebra generated by a pair of two non-commuting smeared operators, $[\hat O_{\bm \gamma}, \hat{O}_{\bm \gamma' }] = i\openone$. The language of modes is particularly useful for describing quasifree (zero mean Gaussian) states---such as the Minkowski vacuum, allowing subsystems, correlations and entanglement to be characterized in terms of the expectation values of the corresponding mode operators. Indeed, the degrees of freedom of a quasifree state are fully encoded in a covariance matrix $\patricianocry = \langle{\{\hat O}_{\bm \gamma},\hat{O}_{\bm \gamma'}\}\rangle$. In particular, restricting $\bm \sigma$ to act on a basis of generators of a local algebra $\mathcal{A}(\Sigma_U)$, one obtains a local representation of the state. In what follows we will use this strategy to represent degrees of freedom associated to disjoint regions $\Sigma_\tc{a}$ and $\Sigma_\tc{b}$ in $\mathcal{A}(\Sigma_\tc{a}\cup \Sigma_\tc{b})\cong \mathcal{A}(\Sigma_\tc{a})\otimes \mathcal{A}(\Sigma_\tc{b})$ (see e.g.~\cite{witten}) and quantify entanglement between them.

\sect{Entanglement between two finite regions}A core concept in our understanding of vacuum entanglement in QFT is the Reeh-Schlieder\footnote{The Reeh-Schlieder theorem states that the Minkowski vacuum is cyclic and separating for any local algebra. That is, acting with local algebra elements on the vacuum can effectively generate any state, and the vacuum does not belong to the kernel of any non-zero local operator.} theorem~\cite{reeh1961bemerkungen,witten}. It is central to prove that the vacuum maximally violates Bell inequalities between two complementary regions~\cite{vacuumBell,summers_bells_1987,vacuumEntanglement}, and it is a core element in showing that there is non-zero vacuum entanglement between \textit{any} two spacelike separated regions (see Cor. 5.1.1 in~\cite{Hollands:2017dov}). Moreover, the mutual information---a measure of the combined classical and quantum correlations---between finite non-overlapping regions is finite and is an upper bound for entanglement measures such as the relative entropy of entanglement~\cite{Hollands:2017dov}. However, neither the Reeh-Schlieder theorem nor the mutual information allow one to explicitly \textit{compute} entanglement between finite regions\footnote{Indeed, the mutual information and entanglement measures can behave drastically differently in QFT~\cite{ABRM_deSitter_short,ABRM_deSitter_long}.}.

Although vacuum entanglement between non-complementary regions in relativistic QFTs has not yet been quantified, recent progress has been made using lattice approximations to massive field theories~\cite{KlcoUVIR,KlcoAllDist2025}. Specifically, Klco, Beck, and Savage (KBS) introduced in~\cite{klco_entanglement_2022} an algorithm for identifying pairs of modes that contribute individually to the entanglement between subregions. This method can be generalized and applied to any bipartite Gaussian quantum system~\cite{partnerformula}, such as the vacuum state restricted to two finite non-overlapping regions.








\newcommand{\lapl}{\xi}

We study vacuum entanglement between disjoint local algebras---associated with spherical regions---of a massless scalar field in 1+3 dimensions. We consider two spherical regions $\Sigma_\tc{a}$ and $\Sigma_\tc{b}$ of radius $R$, whose centers are separated by $\bm L = (0,0,L)$, so that the spheres' separation is $d = L-2R$ with $L>2R$. We parametrize the generators of the corresponding local algebras $\mathcal{A}(\Sigma_\tc{a})$ and $\mathcal{A}(\Sigma_\tc{b})$ in terms of the $L^2$ normalized real eigenfunctions of the Laplacian with Dirichlet boundary conditions. We denote these functions by $\lapl_{\tc{a},\alpha}(\vec{x})$ and $\lapl_{\tc{b},\alpha}(\vec{x})$, with $\alpha = (n,l,m)$ corresponding to the usual spherical quantum numbers $1\leq n < \infty$, $0\leq l< \infty$ and $-l\leq m\leq l$. The generators of the local algebras can be written as $(\hat{\Phi}(F_{\tc{a},\alpha}), \hat{\Pi}(G_{\tc{a},\beta}))$ and $(\hat{\Phi}(F_{\tc{b},\alpha}), \hat{\Pi}(G_{\tc{b},\beta}))$, where $F_{\tc{a},\alpha}(\vec{x}) = \,\lapl_{\tc{a},\alpha}(\vec{x})/\sqrt{R}$, $G_{\tc{a},\beta}(\vec{x}) = \sqrt{R}\lapl_{\tc{a},\beta}(\vec{x})$, and analogous for $\Sigma_\tc{b}$ (see Appendix~\ref{app:calculations} for explicit expressions). 

In this setup, the vacuum properties between the two regions are fully encoded in expected values of product of pairs of these generators. However, it is not physically possible to access these infinitely many degrees of freedom: any interaction or measurement of a quantum field naturally has a finite UV cutoff. In a local representation of field modes, this UV cutoff is introduced by restricting the description to a finite number of modes $N$, which we will assume from this point on. With this assumption, we can then construct a finite dimensional covariance matrix $\bm \sigma$ that represents the local vacuum degrees of freedom:
\begin{equation}
    \bm \sigma = \left(\begin{matrix}
        \bm \sigma_{\tc{aa}} & \bm \sigma_{\tc{ab}} \\
        \bm \sigma_{\tc{ba}} & \bm \sigma_{\tc{bb}}
    \end{matrix}\right),
\end{equation}
where for $\tc{I},\tc{J}\in\{\tc{A},\tc{B}\}$ we have $ \bm \sigma_\tc{ij} = \bm \sigma_{\Phi_\tc{i}\Phi_\tc{j}} \oplus \bm \sigma_{\Pi_\tc{i}\Pi_\tc{j}}$, and $\sigma_{\Phi_\tc{i}\Phi_\tc{j}}$, $\sigma_{\Pi_\tc{i}\Pi_\tc{j}}$ are $N\times N$ matrices with components
\begin{equation}\label{eq:cov}
    \begin{gathered}
    (\bm \sigma_{\Phi_\tc{i}\Phi_\tc{j}})_{\alpha \beta} = \langle\{\hat{\Phi}(F_{\tc{i},\alpha}),\hat{\Phi}(F_{\tc{j},\beta})\}\rangle,\\
    (\bm \sigma_{\Pi_\tc{i}\Pi_\tc{j}})_{\alpha \beta} = \langle\{\hat{\Pi}(G_{\tc{i},\alpha}),\hat{\Pi}(G_{\tc{j},\beta})\}\rangle.
    \end{gathered}
\end{equation}
The covariance matrix $\bm \sigma$ determines the Gaussian Wigner function of the vacuum within the subspace spanned by these modes. It is symmetric and---denoting the symplectic form by $\bm \Omega$---it satisfies $\bm \sigma \geq \ii \bm \Omega$: its symplectic eigenvalues are larger than one, encoding the positivity of the state~\cite{serafini2017quantum}.

Within this Gaussian description, one can verify whether a system is entangled using the Peres-Horodecki criterion~\cite{peresCritOG,horodeckiCritOG,gaussEnt}. Unlike for the physical covariance matrix $\bm \sigma$, if $A$ and $B$ are entangled, the symplectic eigenvalues of the partial transposed covariance matrix $\bm \sigma^{\Gamma_\tc{a}}$, $\tilde{\nu}_k$, might be smaller than one. This motivates the definition of an entanglement monotone, the logarithmic negativity (LN),
\begin{equation}
    E_{\mathcal{N}} = \sum_{k} \mathrm{max}\{0,-\log\tilde\nu_{k}\}.
\end{equation}
The LN is an upper bound of distillable entanglement~\cite{VidalNegativity,plenio,gaussEnt}, and has the operational interpretation of the entanglement cost under quantum operations preserving the positivity of the partial transpose~\cite{PhysRevLett.90.027901}.



The KBS algorithm~\cite{klco_entanglement_2022} selects the individual modes associated with each symplectic eigenvalue of $\bm \sigma^{\Gamma_A}$ smaller than one. This procedure can be understood as an explicit realization of the entanglement partner construction for Gaussian mixed states developed  in Refs.~\cite{partnerformula,partnersqft}, where the existence and uniqueness of such partners are established. For a bipartite Gaussian system $AB$, the entanglement partner\footnote{In terms of the covariance matrix, the entanglement partner of $A$ corresponds to the projection of the subspace spanned by the eigenvectors of $\bm \sigma^{\Gamma_\tc{a}}$ with symplectic eigenvalues smaller than one into the subspace of $B$.} of a subsystem $A$ is the subsystem of $B$ that encodes all distillable entanglement shared with $A$. This allows us to identify the pair of modes $(\hat{\Phi}(\memf_\tc{a}), \hat{\Pi}(\memg_\tc{a}))$ and $(\hat{\Phi}(\memf_\tc{b}), \hat{\Pi}(\memg_\tc{b}))$---one supported in $\Sigma_\tc{a}$ and another in $\Sigma_\tc{b}$, respectively---that contribute the most to the total value of the LN. We refer to this pair as the most entangled modes (MEMs).

\begin{figure}[h!]
\centering
\begin{tikzpicture}
    \node at (0,0) {\includegraphics[width = 8.6cm]{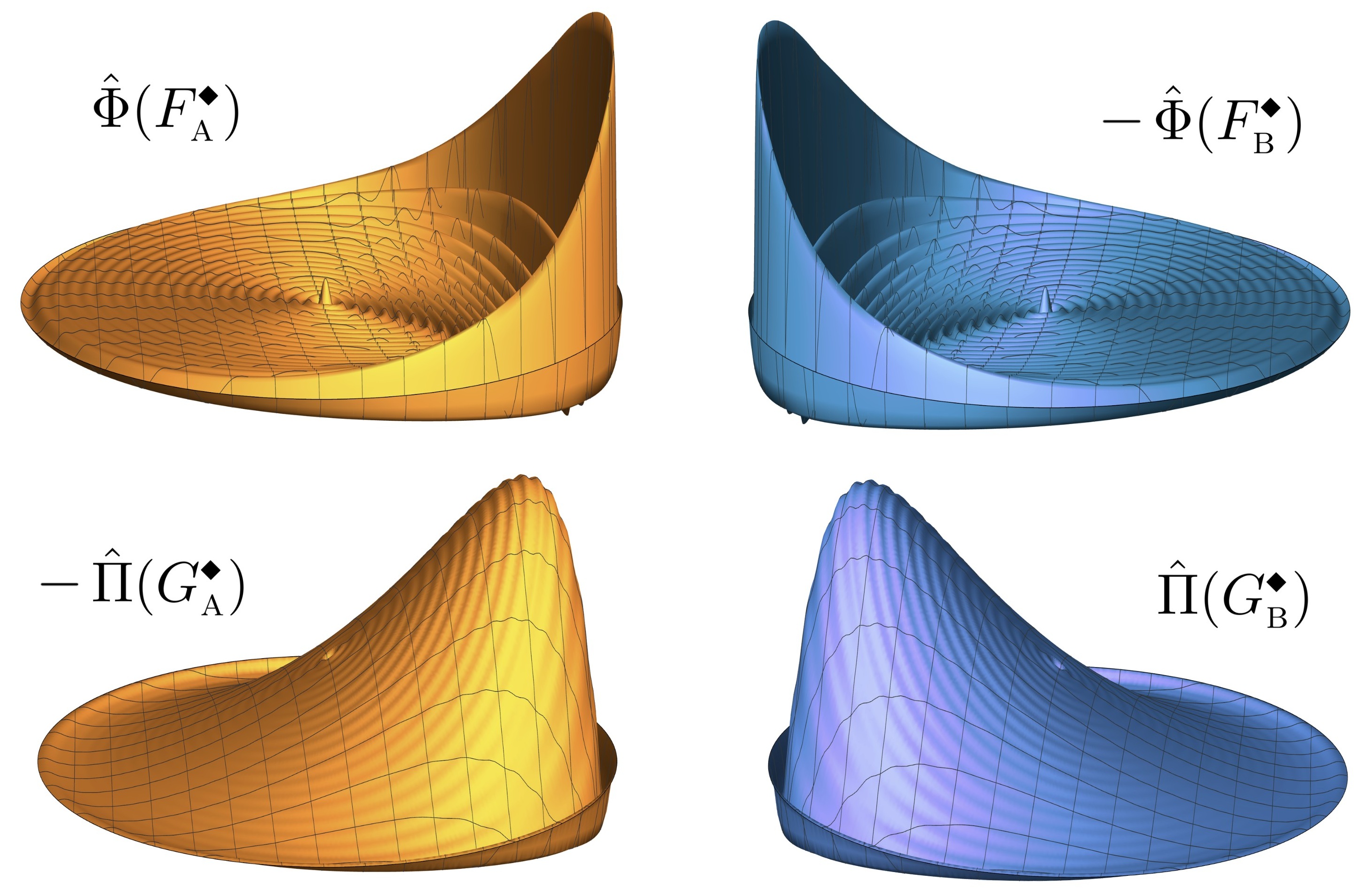}}; 
    \draw[<->] (-2.3,-2.815) -- (-0.34,-2.815); 
    \draw[<->] (-0.36,-2) -- (0.47,-2); 
    \node at (-1.3,-3.03) {\footnotesize{$R$}};
    \node at (0.07,-1.6) {$d \!= \!\frac{R}{2}$};
\end{tikzpicture}
\caption{Depiction of  the amplitude and momentum spatial smearings corresponding to the MEMs  $(\hat{\Phi}(\memf_\tc{a}), \hat{\Pi}(\memg_\tc{a}))$ and $(\hat{\Phi}(\memf_\tc{b}), \hat{\Pi}(\memg_\tc{b}))$ between two spheres of radius $R$ separated by a distance of $d = R/2$ in the $xz$ plane. The modes were studied in the subspace $1\leq n\leq 40$ and $0\leq l\leq 30$.
}
\label{fig:modes}
\end{figure}

\begin{figure*}[t!] 
\begin{tikzpicture}
    \node at (0,0) {\includegraphics[width = \textwidth]{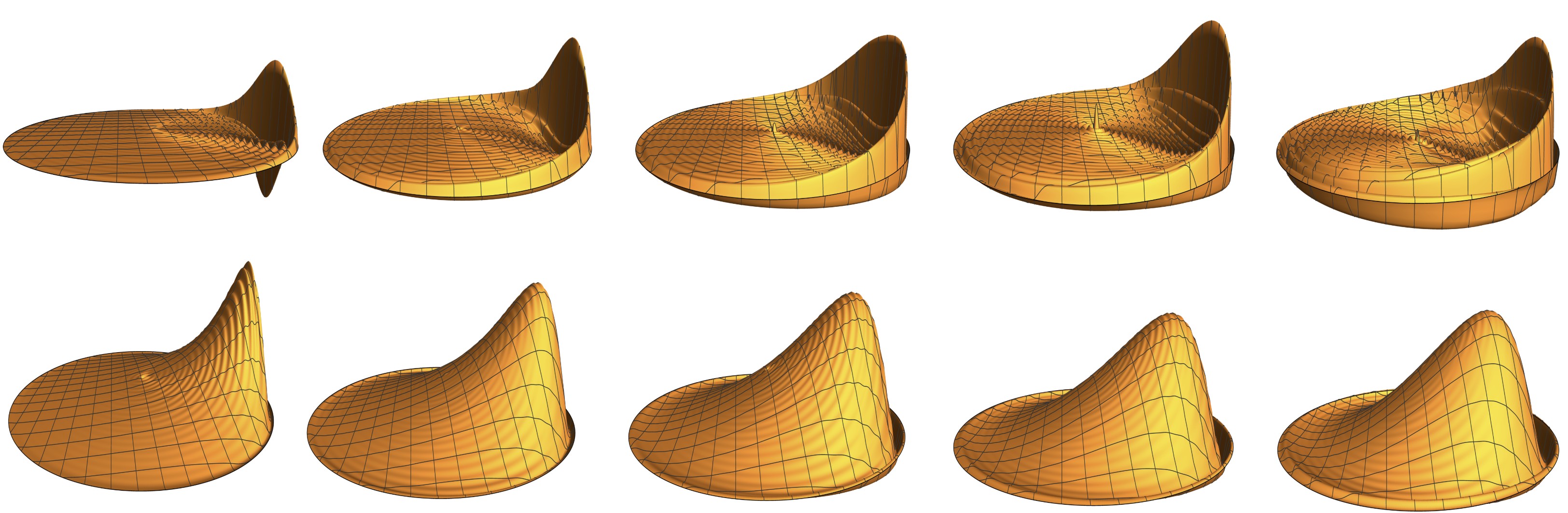}}; 
    \node at (-7.3,3) {$d \!= 0.1\,R$};
    \node at (-3.7,3) {$d \!= 0.3\,R$};
    \node at (-0.1,3) {$d \!= 0.5\,R$};
    \node at (3.5,3) {$d \!= 0.7\,R$};
    \node at (7.3,3) {$d \!= 0.9\,R$};
    \node at (-8.4,2.2) {$\hat{\Phi}(\memf_\tc{a})$};
    \node at (-8.4,-0.65) {$-\hat{\Pi}(\memg_\tc{a})$};
\end{tikzpicture}
\caption{Depiction of the field amplitude and momentum spatial smearings corresponding to the MEMs  $(\hat{\Phi}(\memf_\tc{a}), \hat{\Pi}(\memg_\tc{a}))$ between two spheres of radius $R$ separated by different distances $d$ in the $xz$ plane. The modes were studied in the subspace $1\leq n\leq 40$ and $0\leq l\leq 30$.}\label{fig:MEMsDist}
\end{figure*}
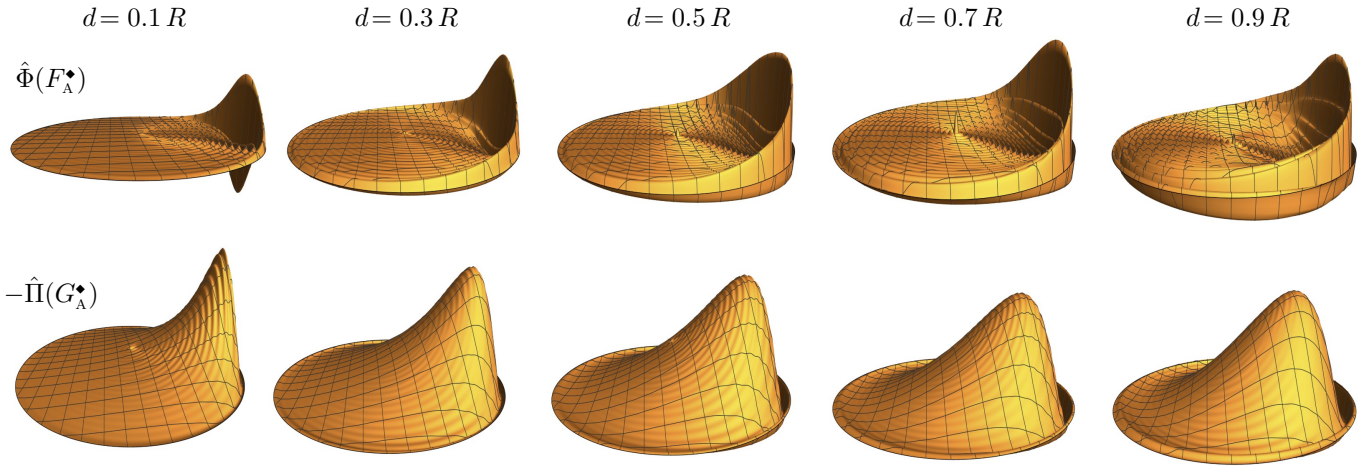

Figure~\ref{fig:modes} shows the spatial profile of the MEMs of a massless scalar field when the spherical regions $\Sigma_A$ and $\Sigma_B$ are separated by a distance of $d=0.5R$, representing the degrees of freedom of $N=1240$ cylindrically symmetric\footnote{The eigenspace corresponding to each symplectic eigenvalue of $\bm \sigma^{\Gamma_A}$ contains a cylindrically symmetric subspace, as a result of the  symmetry of the problem.   } modes in each region, corresponding to $1\leq n\leq 40$ and $0\leq l\leq 30$ with $m=0$. We find that these MEMs are entangled with $E_{\mathcal{N}} \approx 2.2\times 10^{-3}$. In Appendix~\ref{app:convergence_mems} we argue that this approximation accurately describes the MEMs momentum profiles. We also argue that the shapes of the amplitude MEMs do not converge to a square integrable function as $N\to\infty$\footnote{In~\cite{Jason}, the authors identify the negativity cores in 1+1D---the modes that individually contribute to the LN between two intervals---and show that they behave as ${\sim h^{-1/2}\sin(\alpha \log(\beta h))}$, where $h$ is the distance to the boundary.}, in agreement with the results of~\cite{Jason}. Even though the amplitude shapes do not converge in this limit, the functions in Fig.~\ref{fig:modes} are the cylindrically symmetric MEMs that can be physically accessed in the subspace $n\leq 40$, $l\leq 30$. The mode functions in regions A and B are sign-flipped mirror images with oscillations due to the cutoff in $n$. $\memf_\tc{a}$ and $\memf_{\tc{b}}$ are highly oscillatory near the boundary, where they peak, suggesting that the cutoff $n\leq 40$ cannot capture the full range of their oscillations. In contrast, the shapes of $\memg_\tc{a}$ and $\memg_{\tc{b}}$ are well-captured within our cutoff, peaking slightly away from the boundary. This behaviour is consistent with previous results for lattice QFTs (see e.g.~\cite{KlcoUVIR,klco_entanglement_2022,KlcoAllDist2025}).


Figure~\ref{fig:MEMsDist} shows the spatial profile of $\memf_\tc{a}$ and $\memg_\tc{a}$ for several separations $d$ between the spheres. When the distance $d$ is small compared to the radius $R$, the functions become highly localized near the points of minimal distance between the spheres. This behaviour can be attributed to the fact that vacuum correlations are mostly localized at the boundary~\cite{kelly}. As the separation increases, both functions become more oscillatory and their peaks become broader, moving away from the boundary toward the center. The observed shift of support away from the boundary suggests that more oscillatory modes extending into the bulk probe long-distance correlations, matching the lattice behaviour found in~\cite{KlcoUVIR}.





\begin{figure}[h!] 
\centering
\includegraphics[width = 8.6cm]{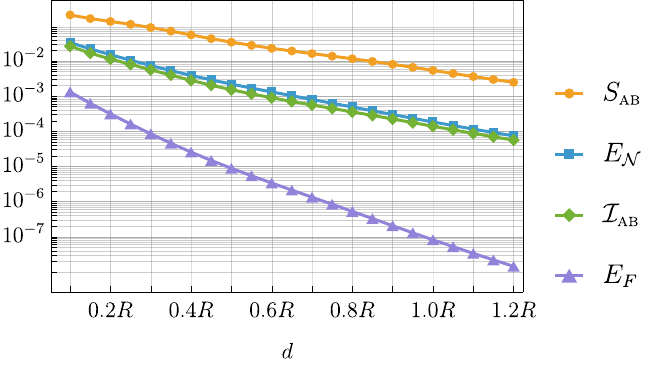}
\caption{Logarithmic negativity $E_{\mathcal{N}}$, Von-Neumann entropy $S_\tc{ab}$, mutual information $\mathcal{I}_\tc{ab}$, and entanglement of formation $E_F$ of MEMs in the subspace $1\leq n\leq 40$ and $0\leq l\leq 30$ as a function of their separation $d$. }\label{fig:NegImpPlot}
\end{figure}

In Figure~\ref{fig:NegImpPlot}, we plot the LN ($E_\mathcal{N}$), the mutual information ($\mathcal{I}_\tc{ab}$), the entanglement of formation ($E_F$)~\cite{gaussForm}, and the Von-Neumann entropy ($S_{\tc{ab}}$) of the MEMs as a function of distance. We observe an exponential decay of both quantum and classical correlations---as quantified by the LN and mutual information, respectively---, as well as impurity (quantified by $S_\tc{ab}$) with distance. Notice that the LN, unlike the entanglement of formation, is not bounded from above by the mutual information. The entanglement of formation is significantly smaller and decays faster with distance when compared to the LN.

\sect{Optimal Entanglement Extraction}Vacuum entanglement extraction has been an active field of research since the 90s~\cite{Valentini1991,Reznik2003,Reznik1}. The topic has been discussed theoretically using operational tools in the protocol of entanglement harvesting~\cite{Nick,NickEdu2014,Pozas-Kerstjens:2015,Pozas2016,HarvestingBHLaura,foo,HarvestingAccelerationRobb,EricksonZero,ericksonNew,carol,boris,quantClass,FullyRelativisticEH}, and more recently experimental proposals have been put forward for measuring vacuum entanglement in both fundamental and effective quantum field theories~\cite{Mo_Settembrini2022VacuumCorrelationsNatCommun,Mo_VacuumEntanglement_BEC_Gooding_2024,adamExperimentalEH2025,BECEH}. However, the protocol of entanglement harvesting is usually studied in the context of perturbation theory, where the entanglement extracted is of second order in the coupling of probes with the field, being restricted to negativities of the order of $10^{-4}$ (see~\cite{marcos} for a detailed discussion).

We can now prescribe a protocol in which probes directly extract the vacuum entanglement present in the MEMs identified in the previous section. This can be done using ideas originally employed in~\cite{kelly,KlcoAllDist2025}, where two harmonic oscillator probes perform mode swap operations with the respective MEMs in each region. Specifically, consider two harmonic oscillators with mass $m$, frequency $\omega$, and dimensionless quadratures $\hat{Q}_\tc{a}, \hat{P}_\tc{a}$, $\hat{Q}_\tc{b}, \hat{P}_\tc{b}$\footnote{The dimensionless quadratures are defined as $\hat{Q}_i = \sqrt{m\omega}\hat{q}_i$ and $\hat{P}_i = \hat{p}_i/\sqrt{m\omega}$ in terms of the position and momentum operators.}, and let $\memf_\tc{a}, \memg_\tc{a}$ and $\memf_\tc{b}$, $\memg_\tc{b}$ be the shape of the amplitude and momentum MEMs. An instantaneous interaction of the form
\begin{align}\label{eq:Hint}
    \hat{H}_{\text{int}}(t) =& \frac{\pi}{2} \delta(t) \left(\hat{Q}_\tc{a}\hat{\Pi}(\memg_\tc{a})  + \hat{P}_\tc{a}\hat{\Phi}(\memf_\tc{a})\right) \\&+\frac{\pi}{2} \delta(t) \left(\,\hat{Q}_\tc{b}\hat{\Pi}(\memg_\tc{b})  +\hat{P}_\tc{b}\hat{\Phi}(\memf_\tc{b})\right) \nonumber
\end{align}
locally couples oscillator A with the field in the sphere $\Sigma_\tc{a}$ and oscillator B with the field in the sphere $\Sigma_\tc{b}$ at $t=0$. This interaction generates a mode swap unitary $\hat{U}_\text{SWAP}$ at time $t = 0$. After the interaction, the final state of the harmonic oscillators will be exactly the same as that of the MEMs, extracting the optimal amount of entanglement from the spherical regions they couple to. For instance, considering a separation between the centers of the spheres of $d = 0.1 R$, one obtains probes with $E_\mathcal{N} \approx 4\times 10^{-2}$, surpassing any value that could be described with leading order perturbation theory in entanglement harvesting.

Although the interaction~\eqref{eq:Hint} allows the implementation of a mode swap operation, it is not immediately clear how it could be implemented in practice. Not only would one require to find probes that can directly couple to the MEMs, but the interaction would also have to be implemented at arbitrarily short time scales. However, the Hamiltonian $\hat{H}_\text{int}$ is not the unique interaction that can generate a mode swap unitary. Evidence from local couplings in QFT suggest that the shape of the wavefunctions of the initial state of the probes and temporal profile of the interaction determine the 4-smeared field observables $\hat{\phi}(f)$ that it couples to~\cite{Unruh-Wald,generalPD,QFTPD}. Using the relationship $\hat{\phi}(f) = - \hat{\Phi}(G) + \hat{\Pi}(F)$ could then be a promising venue for developing probes that can directly couple to vacuum MEMs. 



\sect{Conclusions}We applied the entanglement partner construction for the vacuum of a 1+3 real scalar field in Minkowski spacetime restricted to two non-overlapping spheres, finding the approximate shape of the most entangled modes (MEMs) between these regions. To obtain a representation of the modes, we chose a basis of cylindrically symmetric functions for the regions and represented $N=1240$ local field degrees of freedom associated with the smeared field and momentum in each region. We then obtained an approximate representation of the MEMs in terms of smeared field and momentum operators in the spheres (see Fig.~\ref{fig:modes}).

As shown in~\cite{KlcoUVIR,klco_entanglement_2022,Jason}, the MEMs contain the majority of the accessible vacuum entanglement between the two regions and pinpoint the local degrees of freedom that encode the quantum correlations between the regions. We also discussed an optimal entanglement extraction protocol using two harmonic oscillator probes that perform a mode swap operation with the MEMs. This method allows a non-perturbative treatment and extracts a non-negligible amount of entanglement from the vacuum. This is particularly relevant for applications such as relativistic quantum computing~\cite{Martin-Martinez_Aasen_Kempf_2013,Martin-Martinez_Sutherland_2014,Layden_Martin-Martinez_Kempf_2016,AsplingLawler2024,phil}, where although vacuum entanglement is a valuable resource, non-perturbative theoretical tools are scarce.


Our results are a first step towards understanding the local entanglement structure of quantum field theory in Minkowski spacetime. The MEMs simplify the study of entanglement between finite regions and define a clear focus for future experimental efforts towards measuring spacelike vacuum entanglement.




\acknowledgements

The authors thank Ivan Agullo, B\'eatrice Bonga, Eduardo Mart\'in-Mart\'inez, Jos\'e Polo-G\'omez, Jorma Louko, Sergi Nadal-Gisbert, Adrià Delhom, Mercedes Mart\'in-Benito and Lu\'is Garay for insightful discussions. The authors also thank Ko Sanders, Jorma Louko, and B\'eatrice Bonga for providing valuable comments on a first draft. TRP is hosted at the Physics Department of Stockholm University through a Wenner-Gren Fellowship and is thankful for partial financial support from the Olle Engkvist Foundation (no.225-0062). PRM is thankful for the  financial support from the Royal Commission for the Exhibition of 1851. The Viking cluster was used during this project, which is a high performance compute facility provided by the University of York. We are grateful for computational support from the University of York, IT Services, and the Research IT team.

\bibliography{references}

@article{generalPD,
  title = {Localized nonrelativistic quantum systems in curved spacetimes: A general characterization of particle detector models},
  author = {Perche, T. Rick},
  journal = {Phys. Rev. D},
  volume = {106},
  issue = {2},
  pages = {025018},
  numpages = {20},
  year = {2022},
  month = {Jul},
  publisher = {American Physical Society},
  doi = {10.1103/PhysRevD.106.025018},
  url = {https://link.aps.org/doi/10.1103/PhysRevD.106.025018}
}

@article{RainerEnt,
author = {Verch, Rainer and Werner, Reinhard F.},
title = {DISTILLABILITY AND POSITIVITY OF PARTIAL TRANSPOSES IN GENERAL QUANTUM FIELD SYSTEMS},
journal = {Reviews in Mathematical Physics},
volume = {17},
number = {05},
pages = {545-576},
year = {2005},
doi = {10.1142/S0129055X05002364},

URL = { 
    
        https://doi.org/10.1142/S0129055X05002364
    
    

}
}

@article{BeyondPertUDW,
  title = {Detectors for probing relativistic quantum physics beyond perturbation theory},
  author = {Brown, Eric G. and Mart\'{\i}n-Mart\'{\i}nez, Eduardo and Menicucci, Nicolas C. and Mann, Robert B.},
  journal = {Phys. Rev. D},
  volume = {87},
  issue = {8},
  pages = {084062},
  numpages = {19},
  year = {2013},
  month = {Apr},
  publisher = {American Physical Society},
  doi = {10.1103/PhysRevD.87.084062},
  url = {https://link.aps.org/doi/10.1103/PhysRevD.87.084062}
}

@Article{BenitoJormaUnruhState,
author={Ju{\'a}rez-Aubry, Benito A.
and Louko, Jorma},
title={Quantum fields during black hole formation: how good an approximation is the {U}nruh state?},
journal={J. High Energy Phys.},
year={2018},
month={May},
day={23},
volume={2018},
number={5},
pages={140},
issn={1029-8479},
doi={10.1007/JHEP05(2018)140},
url={https://doi.org/10.1007/JHEP05(2018)140}
}

@article{QFTPD,
  title = {Particle detectors from localized quantum field theories},
  author = {Perche, T. Rick and Polo-G\'omez, Jos\'e and Torres, Bruno de S. L. and Mart\'{\i}n-Mart\'{\i}nez, Eduardo},
  journal = {Phys. Rev. D},
  volume = {109},
  issue = {4},
  pages = {045013},
  numpages = {15},
  year = {2024},
  month = {Feb},
  publisher = {American Physical Society},
  doi = {10.1103/PhysRevD.109.045013},
  url = {https://link.aps.org/doi/10.1103/PhysRevD.109.045013}
}

@article{HarvestingAccelerationRobb,
  title = {Does acceleration assist entanglement harvesting?},
  author = {Liu, Zhihong and Zhang, Jialin and Mann, Robert B. and Yu, Hongwei},
  journal = {Phys. Rev. D},
  volume = {105},
  issue = {8},
  pages = {085012},
  numpages = {9},
  year = {2022},
  month = {Apr},
  publisher = {American Physical Society},
  doi = {10.1103/PhysRevD.105.085012},
  url = {https://link.aps.org/doi/10.1103/PhysRevD.105.085012}
}

@article{VidalNegativity,
  title = {Computable measure of entanglement},
  author = {Vidal, G. and Werner, R. F.},
  journal = {Phys. Rev. A},
  volume = {65},
  issue = {3},
  pages = {032314},
  numpages = {11},
  year = {2002},
  month = {Feb},
  publisher = {American Physical Society},
  doi = {10.1103/PhysRevA.65.032314},
  url = {https://link.aps.org/doi/10.1103/PhysRevA.65.032314}
}

@article{ericksonNew,
  title = {When entanglement harvesting is not really harvesting},
  author = {Tjoa, Erickson and Mart\'{\i}n-Mart\'{\i}nez, Eduardo},
  journal = {Phys. Rev. D},
  volume = {104},
  issue = {12},
  pages = {125005},
  numpages = {21},
  year = {2021},
  month = {Dec},
  publisher = {American Physical Society},
  doi = {10.1103/PhysRevD.104.125005},
  url = {https://link.aps.org/doi/10.1103/PhysRevD.104.125005}
}

@article{boris,
  title = {Harvesting entanglement from the gravitational vacuum},
  author = {Perche, T. Rick and Ragula, Boris and Mart\'{\i}n-Mart\'{\i}nez, Eduardo},
  journal = {Phys. Rev. D},
  volume = {108},
  issue = {8},
  pages = {085025},
  numpages = {58},
  year = {2023},
  month = {Oct},
  publisher = {American Physical Society},
  doi = {10.1103/PhysRevD.108.085025},
  url = {https://link.aps.org/doi/10.1103/PhysRevD.108.085025}
}

@article{vacuumEntanglement,
title = {{The vacuum violates Bell's inequalities}},
journal = {Phys. lett., A},
volume = {110},
number = {5},
pages = {257-259},
year = {1985},
issn = {0375-9601},
doi = {https://doi.org/10.1016/0375-9601(85)90093-3},
url = {https://www.sciencedirect.com/science/article/pii/0375960185900933},
author = {Stephen J. Summers and Reinhard Werner}
}

@article{vacuumBell,
author = {Summers,Stephen J.  and Werner,Reinhard },
title = {Bell’s inequalities and quantum field theory. I. General setting},
journal = {Journal of Mathematical Physics},
volume = {28},
number = {10},
pages = {2440-2447},
year = {1987},
doi = {10.1063/1.527733},
URL = {https://doi.org/10.1063/1.527733}
}

@article{foo,
  title = {Entanglement amplification between superposed detectors in flat and curved spacetimes},
  author = {Foo, Joshua and Mann, Robert B. and Zych, Magdalena},
  journal = {Phys. Rev. D},
  volume = {103},
  issue = {6},
  pages = {065013},
  numpages = {19},
  year = {2021},
  month = {Mar},
  publisher = {American Physical Society},
  doi = {10.1103/PhysRevD.103.065013},
  url = {https://link.aps.org/doi/10.1103/PhysRevD.103.065013}
}

@Article{ericksonBH,
author={Tjoa, Erickson
and Mann, Robert B.},
title={Harvesting correlations in Schwarzschild and collapsing shell spacetimes},
journal={Jour. High Energy Phys.},
year={2020},
month={Aug},
day={28},
volume={2020},
number={8},
pages={155},
issn={1029-8479},
doi={10.1007/JHEP08(2020)155},
url={https://doi.org/10.1007/JHEP08(2020)155}
}

@Article{Cong2019,
author={Cong, Wan
and Tjoa, Erickson
and Mann, Robert B.},
title={Entanglement harvesting with moving mirrors},
journal={J. High Energy Phys.},
year={2019},
month={Jun},
day={07},
volume={2019},
number={6},
pages={21},
issn={1029-8479},
doi={10.1007/JHEP06(2019)021},
url={https://doi.org/10.1007/JHEP06(2019)021}
}

@article{Pozas2016,
  title = {Entanglement harvesting from the electromagnetic vacuum with hydrogenlike atoms},
  author = {Pozas-Kerstjens, Alejandro and Mart\'{i}n-Mart\'{i}nez, Eduardo},
  journal = {Phys. Rev. D},
  volume = {94},
  issue = {6},
  pages = {064074},
  numpages = {27},
  year = {2016},
  month = {Sep},
  publisher = {American Physical Society},
  doi = {10.1103/PhysRevD.94.064074},
  url = {https://link.aps.org/doi/10.1103/PhysRevD.94.064074}
}

@article{Nick,
	Author = {Greg VerSteeg and Nicolas C. Menicucci},
	Journal = {Phys. Rev. D},
	Pages = {044027},
	Title = {Entangling power of an expanding universe},
	Volume = {79},
	Year = {2009}}

@article{NickEdu2014,
	doi = {10.1088/0264-9381/31/21/214001},
	url = {https://doi.org/10.1088/0264-9381/31/21/214001},
	year = 2014,
	month = {oct},
	publisher = {{IOP} Publishing},
	volume = {31},
	number = {21},
	pages = {214001},
	author = {Eduardo Mart{\'{\i}}n-Mart{\'{\i}}nez and Nicolas C Menicucci},
	title = {Entanglement in curved spacetimes and cosmology},
	journal = {Class. Quantum Gravity}
}

@article{Pozas-Kerstjens:2015,
	Author = {Pozas-Kerstjens, Alejandro and Mart\'{i}n-Mart\'{i}nez, Eduardo},
	Doi = {10.1103/PhysRevD.92.064042},
	Issue = {6},
	Journal = {Phys. Rev. D},
	Month = {Sep},
	Numpages = {18},
	Pages = {064042},
	Publisher = {American Physical Society},
	Title = {Harvesting correlations from the quantum vacuum},
	Url = {http://link.aps.org/doi/10.1103/PhysRevD.92.064042},
	Volume = {92},
	Year = {2015}}

@article{reznik2,
  title = {Long-range entanglement in the {D}irac vacuum},
  author = {Silman, J. and Reznik, B.},
  journal = {Phys. Rev. A},
  volume = {75},
  issue = {5},
  pages = {052307},
  numpages = {5},
  year = {2007},
  month = {May},
  publisher = {American Physical Society},
  doi = {10.1103/PhysRevA.75.052307},
  url = {https://link.aps.org/doi/10.1103/PhysRevA.75.052307}
}

@article{HarvestingBHLaura,
	doi = {10.1088/1361-6382/aae27e},
	url = {https://doi.org/10.1088%2F1361-6382%2Faae27e},
	year = 2018,
	month = {oct},
	publisher = {{IOP} Publishing},
	volume = {35},
	number = {21},
	pages = {21LT02},
	author = {Laura J Henderson and Robie A Hennigar and Robert B Mann and Alexander R H Smith and Jialin Zhang},
	title = {Harvesting entanglement from the black hole vacuum},
	journal = {Class. Quantum Gravity}
}

@article{Reznik1,
	Author = {Benni Reznik and Alex Retzker and Jonathan Silman},
	Eid = {042104},
	Journal = {Phys. Rev. A},
	Number = {4},
	Numpages = {4},
	Pages = {042104},
	Publisher = {APS},
	Title = {{Violating Bell's inequalities in vacuum}},
	Url = {http://link.aps.org/abstract/PRA/v71/e042104},
	Volume = {71},
	Year = {2005}}

@article{Valentini1991,
	Author = {Antony Valentini},
	Doi = {http://dx.doi.org/10.1016/0375-9601(91)90952-5},
	Issn = {0375-9601},
	Journal = {Phys. Lett. A},
	Number = {6-7},
	Pages = {321 - 325},
	Title = {Non-local correlations in quantum electrodynamics},
	Url = {http://www.sciencedirect.com/science/article/pii/0375960191909525},
	Volume = {153},
	Year = {1991}}

@article{Farming,
	Author = {Mart\'{i}n-Mart\'{i}nez, Eduardo and Brown, Eric G. and Donnelly, William and Kempf, Achim},
	Doi = {10.1103/PhysRevA.88.052310},
	Issue = {5},
	Journal = {Phys. Rev. A},
	Month = {Nov},
	Numpages = {15},
	Pages = {052310},
	Publisher = {American Physical Society},
	Title = {Sustainable entanglement production from a quantum field},
	Url = {http://link.aps.org/doi/10.1103/PhysRevA.88.052310},
	Volume = {88},
	Year = {2013}}

@article{Reznik2003,
	Author = {Reznik, Benni},
	Doi = {10.1023/A:1022875910744},
	Issn = {0015-9018},
	Journal = {Foundations of Physics},
	Language = {English},
	Number = {1},
	Pages = {167-176},
	Publisher = {Kluwer Academic Publishers-Plenum Publishers},
	Title = {Entanglement from the Vacuum},
	Url = {http://dx.doi.org/10.1023/A%3A1022875910744},
	Volume = {33},
	Year = {2003}}

@article{witten,
  title = {APS Medal for Exceptional Achievement in Research: Invited article on entanglement properties of quantum field theory},
  author = {Witten, Edward},
  journal = {Rev. Mod. Phys.},
  volume = {90},
  issue = {4},
  pages = {045003},
  numpages = {38},
  year = {2018},
  month = {Oct},
  publisher = {American Physical Society},
  doi = {10.1103/RevModPhys.90.045003},
  url = {https://link.aps.org/doi/10.1103/RevModPhys.90.045003}
}

@misc{AsplingLawler2024, title={Universal Quantum Computing with Field-Mediated {Unruh--DeWitt} Qubits}, url={http://arxiv.org/abs/2402.10173}, note={arXiv:2402.10173}, number={arXiv:2402.10173}, publisher={arXiv}, author={Aspling, Eric and Lawler, Michael}, year={2024}}

@article{Layden_Martin-Martinez_Kempf_2016, title={Universal scheme for indirect quantum control}, volume={93}, ISSN={2469-9926, 2469-9934}, DOI={10.1103/PhysRevA.93.040301}, number={4}, journal={Phys. Rev. A}, author={Layden, David and Martin-Martinez, Eduardo and Kempf, Achim}, year={2016}, pages={040301} }

@misc{Jason,
      title={The negativity core of a {1+1D} massless scalar quantum field}, 
      author={Jason Pye and Atharva Hingane and Robert H. Jonsson},
      year={2026},
      eprint={2605.23824},
      archivePrefix={arXiv},
      primaryClass={hep-th},
      url={https://arxiv.org/abs/2605.23824}, 
}

@article{gaussForm,
  title = {Quantifying entanglement in two-mode {Gaussian} states},
  author = {Tserkis, Spyros and Ralph, Timothy C.},
  journal = {Phys. Rev. A},
  volume = {96},
  issue = {6},
  pages = {062338},
  numpages = {6},
  year = {2017},
  month = {Dec},
  publisher = {American Physical Society},
  doi = {10.1103/PhysRevA.96.062338},
  url = {https://link.aps.org/doi/10.1103/PhysRevA.96.062338}
}

@article{EduAndIvan,
  title = {Multimode nature of spacetime entanglement in QFT},
  author = {Agullo, Ivan and Bonga, B\'eatrice and Mart\'{\i}n-Mart\'{\i}nez, Eduardo and Nadal-Gisbert, Sergi and Perche, T. Rick and Polo-G\'omez, Jos\'e and Ribes-Metidieri, Patricia and Torres, Bruno de S. L.},
  journal = {Phys. Rev. D},
  volume = {111},
  issue = {8},
  pages = {085013},
  numpages = {23},
  year = {2025},
  month = {Apr},
  publisher = {American Physical Society},
  doi = {10.1103/PhysRevD.111.085013},
  url = {https://link.aps.org/doi/10.1103/PhysRevD.111.085013}
}

@article{horodeckiCritOG,
title = {Separability criterion and inseparable mixed states with positive partial transposition},
journal = {Physics Letters A},
volume = {232},
number = {5},
pages = {333-339},
year = {1997},
issn = {0375-9601},
doi = {https://doi.org/10.1016/S0375-9601(97)00416-7},
url = {https://www.sciencedirect.com/science/article/pii/S0375960197004167},
author = {Pawel Horodecki}
}

@article{peresCritOG,
  title = {Separability Criterion for Density Matrices},
  author = {Peres, Asher},
  journal = {Phys. Rev. Lett.},
  volume = {77},
  issue = {8},
  pages = {1413--1415},
  numpages = {0},
  year = {1996},
  month = {Aug},
  publisher = {American Physical Society},
  doi = {10.1103/PhysRevLett.77.1413},
  url = {https://link.aps.org/doi/10.1103/PhysRevLett.77.1413}
}

@article{plenio,
  title = {Logarithmic Negativity: A Full Entanglement Monotone That is not Convex},
  author = {Plenio, M. B.},
  journal = {Phys. Rev. Lett.},
  volume = {95},
  issue = {9},
  pages = {090503},
  numpages = {4},
  year = {2005},
  month = {Aug},
  publisher = {American Physical Society},
  doi = {10.1103/PhysRevLett.95.090503},
  url = {https://link.aps.org/doi/10.1103/PhysRevLett.95.090503}
}

@article{gaussEnt,
  title = {Peres-{Horodecki} Separability Criterion for Continuous Variable Systems},
  author = {Simon, R.},
  journal = {Phys. Rev. Lett.},
  volume = {84},
  issue = {12},
  pages = {2726--2729},
  numpages = {0},
  year = {2000},
  month = {Mar},
  publisher = {American Physical Society},
  doi = {10.1103/PhysRevLett.84.2726},
  url = {https://link.aps.org/doi/10.1103/PhysRevLett.84.2726}
}

@article{Martin-Martinez_Sutherland_2014, title={Quantum gates via relativistic remote control}, volume={739}, ISSN={0370-2693}, DOI={10.1016/j.physletb.2014.10.038}, journal={Phys. Lett. B}, author={Martín-Martínez, Eduardo and Sutherland, Chris}, year={2014}, pages={74–82} }

@article{Mo_Onoe2022RapidlySwitchedUDW,
  title   = {Realizing a rapidly switched {Unruh-DeWitt} detector through electro-optic sampling of the electromagnetic vacuum},
  author  = {Onoe, Sho and Guedes, Thiago L. M. and Moskalenko, Andrey S. and Leitenstorfer, Alfred and Burkard, Guido and Ralph, Timothy C.},
  journal = {Physical Review D},
  volume  = {105},
  number  = {5},
  pages   = {056023},
  year    = {2022},
  doi     = {10.1103/PhysRevD.105.056023}
}

@article{Martin-Martinez_Aasen_Kempf_2013, title={Processing quantum information with relativistic motion of atoms}, volume={110}, ISSN={0031-9007, 1079-7114}, DOI={10.1103/PhysRevLett.110.160501}, number={16}, journal={Phys. Rev. Lett.}, author={Martin-Martinez, Eduardo and Aasen, David and Kempf, Achim}, year={2013}, pages={160501} }

@article{phil,
  title = {Universal Quantum Computer from Relativistic Motion},
  author = {LeMaitre, Philip A. and Perche, T. Rick and Krumm, Marius and Briegel, Hans J.},
  journal = {Phys. Rev. Lett.},
  volume = {134},
  issue = {19},
  pages = {190601},
  numpages = {7},
  year = {2025},
  month = {May},
  publisher = {American Physical Society},
  doi = {10.1103/PhysRevLett.134.190601},
  url = {https://link.aps.org/doi/10.1103/PhysRevLett.134.190601}
}

@article{reznikLattice,
  title = {Critical and noncritical long-range entanglement in {Klein-Gordon} fields},
  author = {Marcovitch, S. and Retzker, A. and Plenio, M. B. and Reznik, B.},
  journal = {Phys. Rev. A},
  volume = {80},
  issue = {1},
  pages = {012325},
  numpages = {4},
  year = {2009},
  month = {Jul},
  publisher = {American Physical Society},
  doi = {10.1103/PhysRevA.80.012325},
  url = {https://link.aps.org/doi/10.1103/PhysRevA.80.012325}
}

@book{Haag,
  doi = {10.1007/978-3-642-97306-2},
  url = {https://doi.org/10.1007/978-3-642-97306-2},
  year = {1992},
  publisher = {Springer-Verlag Berlin Heidelberg},
  author = {Rudolf Haag},
  title = {Local Quantum Physics: Fields, Particles, Algebras}
}

@article{Unruh-Wald,
	Author = {Unruh, William G. and Wald, Robert M.},
	Doi = {10.1103/PhysRevD.29.1047},
	Issue = {6},
	Journal = {Phys. Rev. D},
	Month = {Mar},
	Pages = {1047--1056},
	Publisher = {American Physical Society},
	Title = {{What happens when an accelerating observer detects a Rindler particle}},
	Volume = {29},
	Year = {1984}
}

@book{Wald2,
author={Robert Manuel Wald},
title={Quantum Field Theory in Curved Spacetime and Black Hole Thermodynamics},
publisher={The University of Chicago Press},
year={1994}
}

@book{gradshteyn,
  author = {Gradshteyn, I. S. and Ryzhik, I. M.},
  description = {MR: Publications results for "MR Number=(2360010)"},
  edition = {{S}eventh},
  isbn = {978-0-12-373637-6; 0-12-373637-4},
  mrclass = {00A22 (33-00 65-00 65A05)},
  mrnumber = {2360010 (2008g:00005)},
  publisher = {Elsevier/Academic Press, Amsterdam},
  title = {Table of integrals, series, and products},
  year = 2007
}

@article{carol,
  title = {Harvesting entanglement from complex scalar and fermionic fields with linearly coupled particle detectors},
  author = {Perche, T. Rick and Lima, Caroline and Mart\'{\i}n-Mart\'{\i}nez, Eduardo},
  journal = {Phys. Rev. D},
  volume = {105},
  issue = {6},
  pages = {065016},
  numpages = {24},
  year = {2022},
  month = {Mar},
  publisher = {American Physical Society},
  doi = {10.1103/PhysRevD.105.065016},
  url = {https://link.aps.org/doi/10.1103/PhysRevD.105.065016}
}

@article{B,
  title = {Spin Entanglement Witness for Quantum Gravity},
  author = {Bose, Sougato and Mazumdar, Anupam and Morley, Gavin W. and Ulbricht, Hendrik and Toro\ifmmode \check{s}\else \v{s}\fi{}, Marko and Paternostro, Mauro and Geraci, Andrew A. and Barker, Peter F. and Kim, M. S. and Milburn, Gerard},
  journal = {Phys. Rev. Lett.},
  volume = {119},
  issue = {24},
  pages = {240401},
  numpages = {6},
  year = {2017},
  month = {Dec},
  publisher = {American Physical Society},
  doi = {10.1103/PhysRevLett.119.240401},
  url = {https://link.aps.org/doi/10.1103/PhysRevLett.119.240401}
}

@article{areaLaw1993,
  title = {Entropy and area},
  author = {Srednicki, Mark},
  journal = {Phys. Rev. Lett.},
  volume = {71},
  issue = {5},
  pages = {666--669},
  numpages = {0},
  year = {1993},
  month = {Aug},
  publisher = {American Physical Society},
  doi = {10.1103/PhysRevLett.71.666},
  url = {https://link.aps.org/doi/10.1103/PhysRevLett.71.666}
}

@article{SorkinArea1986,
  title = {Quantum source of entropy for black holes},
  author = {Bombelli, Luca and Koul, Rabinder K. and Lee, Joohan and Sorkin, Rafael D.},
  journal = {Phys. Rev. D},
  volume = {34},
  issue = {2},
  pages = {373--383},
  numpages = {0},
  year = {1986},
  month = {Jul},
  publisher = {American Physical Society},
  doi = {10.1103/PhysRevD.34.373},
  url = {https://link.aps.org/doi/10.1103/PhysRevD.34.373}
}

@Article{kelly,
author={de S. L. Torres, Bruno
and Wurtz, Kelly
and Polo-G{\'o}mez, Jos{\'e}
and Mart{\'i}n-Mart{\'i}nez, Eduardo},
title={Entanglement structure of quantum fields through local probes},
journal={J. High Energy Phys.},
year={2023},
month={May},
day={08},
volume={2023},
number={5},
pages={58},
issn={1029-8479},
doi={10.1007/JHEP05(2023)058},
url={https://doi.org/10.1007/JHEP05(2023)058}
}

@article{derivativeJorma,
doi = {10.1088/0264-9381/31/24/245007},
url = {https://dx.doi.org/10.1088/0264-9381/31/24/245007},
year = {2014},
month = {nov},
publisher = {IOP Publishing},
volume = {31},
number = {24},
pages = {245007},
author = {Benito A Juárez-Aubry and Jorma Louko},
title = {{Onset and decay of the 1 + 1 Hawking–Unruh effect: what the derivative-coupling detector saw}},
journal = {Class. Quantum Gravity}
}

@article{quantClass,
  title = {Role of quantum degrees of freedom of relativistic fields in quantum information protocols},
  author = {Perche, T. Rick and Mart\'{\i}n-Mart\'{\i}nez, Eduardo},
  journal = {Phys. Rev. A},
  volume = {107},
  issue = {4},
  pages = {042612},
  numpages = {20},
  year = {2023},
  month = {Apr},
  publisher = {American Physical Society},
  doi = {10.1103/PhysRevA.107.042612},
  url = {https://link.aps.org/doi/10.1103/PhysRevA.107.042612}
}

@article{EricksonZero,
  title = {Vacuum entanglement harvesting with a zero mode},
  author = {Tjoa, Erickson and Mart\'{\i}n-Mart\'{\i}nez, Eduardo},
  journal = {Phys. Rev. D},
  volume = {101},
  issue = {12},
  pages = {125020},
  numpages = {10},
  year = {2020},
  month = {Jun},
  publisher = {American Physical Society},
  doi = {10.1103/PhysRevD.101.125020},
  url = {https://link.aps.org/doi/10.1103/PhysRevD.101.125020}
}

@article{max,
doi = {10.1088/1361-6382/ac1b08},
url = {https://dx.doi.org/10.1088/1361-6382/ac1b08},
year = {2021},
month = {sep},
publisher = {IOP Publishing},
volume = {38},
number = {19},
pages = {195029},
author = {Maximilian H Ruep},
title = {Weakly coupled local particle detectors cannot harvest entanglement},
journal = {Class. Quantum Gravity}
}

@article{ABRM_ubiquitous,
    author = "Agullo, Ivan and Bonga, B{\'e}atrice and Ribes-Metidieri, Patricia and Kranas, Dimitrios and Nadal-Gisbert, Sergi",
    title = "{How ubiquitous is entanglement in quantum field theory?}",
    eprint = "2302.13742",
    archivePrefix = "arXiv",
    primaryClass = "quant-ph",
    doi = "10.1103/PhysRevD.108.085005",
    journal = "Phys. Rev. D",
    volume = "108",
    number = "8",
    pages = "085005",
    year = "2023"
}

@article{FullyRelativisticEH,
  title = {Fully relativistic entanglement harvesting},
  author = {Perche, T. Rick and Polo-G\'omez, Jos\'e and Torres, Bruno de S. L. and Mart\'{\i}n-Mart\'{\i}nez, Eduardo},
  journal = {Phys. Rev. D},
  volume = {109},
  issue = {4},
  pages = {045018},
  numpages = {17},
  year = {2024},
  month = {Feb},
  publisher = {American Physical Society},
  doi = {10.1103/PhysRevD.109.045018},
  url = {https://link.aps.org/doi/10.1103/PhysRevD.109.045018}
}

@article{reeh1961bemerkungen,
  title={{Bemerkungen zur Unit{\"a}r{\"a}quivalenz von Lorentzinvarianten Feldern}},
  author={Reeh, Helmut and Schlieder, Siegfried},
  journal={Il Nuovo Cimento (1955-1965)},
  volume={22},
  number={5},
  pages={1051--1068},
  year={1961},
  publisher={Springer}
}

@book{Hollands:2017dov,
    author = {Hollands, Stefan and Sanders, Ko},
    title = {Entanglement Measures and Their Properties in Quantum Field Theory},
    publisher = {Springer Cham},
    year = {2018},
    edition={1},
    series = {SpringerBriefs in Mathematical Physics},
    ISBN ={978-3-319-94901-7}
}

@article{partnerformula,
    author = "Agullo, Ivan and Mart{\'\i}n-Mart{\'\i}nez, Eduardo and Nadal-Gisbert, Sergi and Ribes-Metidieri, Patricia and Yamaguchi, Koji",
    title = "{Correlation and Entanglement partners in {Gaussian} systems}",
    eprint = "2512.11055",
    archivePrefix = "arXiv",
    primaryClass = "quant-ph",
    month = "12",
    year = "2025",
    journal=""
}

@article{partnersqft,
    author = "Agullo, Ivan and Mart{\'\i}n-Mart{\'\i}nez, Eduardo and Nadal-Gisbert, Sergi and Ribes-Metidieri, Patricia and Yamaguchi, Koji",
    title = "{Correlation and Entanglement partners in free quantum field theory }",
    journal="(In preparation)"
    }

@book{serafini2017quantum,
  title={Quantum Continuous Variables: A Primer of Theoretical Methods},
  author={Serafini, A.},
  isbn={9781482246346},
  lccn={2016058596},
 year={2017},
  publisher={CRC Press, Taylor \& Francis Group}
}

@article{PhysRevLett.90.027901,
  title = {Entanglement Cost under Positive-Partial-Transpose-Preserving Operations},
  author = {Audenaert, K. and Plenio, M. B. and Eisert, J.},
  journal = {Phys. Rev. Lett.},
  volume = {90},
  issue = {2},
  pages = {027901},
  numpages = {4},
  year = {2003},
  month = {Jan},
  publisher = {American Physical Society},
  doi = {10.1103/PhysRevLett.90.027901},
  url = {https://link.aps.org/doi/10.1103/PhysRevLett.90.027901}
}

@article{ABRM_deSitter_long,
    author = "Ribes-Metidieri, Patricia and Agullo, Ivan and Bonga, B{\'e}atrice",
    title = "{Entanglement and correlations between local observables in de Sitter spacetime}",
    eprint = "2511.17382",
    archivePrefix = "arXiv",
    primaryClass = "gr-qc",
    doi = "10.1103/qqjv-rv8t",
    journal = "Phys. Rev. D",
    volume = "113",
    number = "6",
    pages = "065001",
    year = "2026"
}

@article{ABRM_deSitter_short,
    author = "Agullo, Ivan and Bonga, B{\'e}atrice and Ribes-Metidieri, Patricia",
    title = "{Inflation does not create entanglement in local observables}",
    doi = "10.1088/1361-6382/ae2a9d",
    journal = "Class. Quant. Grav.",
    volume = "43",
    number = "1",
    pages = "01LT01",
    year = "2026"
}

@article{Mo_VacuumEntanglement_BEC_Gooding_2024,
doi = {10.1088/1367-2630/ad8675},
url = {https://doi.org/10.1088/1367-2630/ad8675},
year = {2024},
month = {oct},
publisher = {IOP Publishing},
volume = {26},
number = {10},
pages = {105001},
author = {Gooding, Cisco and Sachs, Allison and Mann, Robert B and Weinfurtner, Silke},
title = {Vacuum entanglement probes for ultra-cold atom systems},
journal = {New Journal of Physics},
}

@article{Mo_Settembrini2022VacuumCorrelationsNatCommun,
  title   = {Detection of quantum-vacuum field correlations outside the light cone},
  author  = {Settembrini, Francesca Fabiana and Lindel, Frieder and Herter, Alexa Marina and Buhmann, Stefan Yoshi and Faist, J{\'e}r{\^o}me},
  journal = {Nature Communications},
  volume  = {13},
  number  = {1},
  pages   = {3383},
  year    = {2022},
  doi     = {10.1038/s41467-022-31081-1},
  url     = {https://doi.org/10.1038/s41467-022-31081-1}
}

@article{KlcoUVIR,
  title = {Entanglement Spheres and a {UV-IR} Connection in Effective Field Theories},
  author = {Klco, Natalie and Savage, Martin J.},
  journal = {Phys. Rev. Lett.},
  volume = {127},
  issue = {21},
  pages = {211602},
  numpages = {7},
  year = {2021},
  month = {Nov},
  publisher = {American Physical Society},
  doi = {10.1103/PhysRevLett.127.211602},
  url = {https://link.aps.org/doi/10.1103/PhysRevLett.127.211602}
}

@article{adamExperimentalEH2025,
  title = {Towards an experimental implementation of entanglement harvesting in superconducting circuits: Effect of detector gap variation on entanglement harvesting},
  author = {Teixid\'o-Bonfill, Adam and Dai, Xi and Lupascu, Adrian and Mart\'{\i}n-Mart\'{\i}nez, Eduardo},
  journal = {Phys. Rev. A},
  volume = {113},
  issue = {4},
  pages = {043732},
  numpages = {36},
  year = {2026},
  month = {Apr},
  publisher = {American Physical Society},
  doi = {10.1103/wv9n-k3jj},
  url = {https://link.aps.org/doi/10.1103/wv9n-k3jj}
}

@misc{BECEH,
      title={Bose polarons as relativistic {{Unruh-DeWitt}} detectors: Entanglement harvesting from {Bose-Einstein} condensates}, 
      author={T. Rick Perche and Francesco Gozzini and Markus K. Oberthaler},
      year={2026},
      eprint={2512.21381},
      archivePrefix={arXiv},
      primaryClass={quant-ph},
      url={https://arxiv.org/abs/2512.21381}, 
}

@article{KlcoAllDist2025,
  title = {Detecting spacelike vacuum entanglement at all distances and promoting negativity to a necessary and sufficient entanglement measure in many-body regimes},
  author = {Gao, Boyu and Klco, Natalie},
  journal = {Phys. Rev. A},
  volume = {112},
  issue = {1},
  pages = {012430},
  numpages = {13},
  year = {2025},
  month = {Jul},
  publisher = {American Physical Society},
  doi = {10.1103/m9w1-ppqz},
  url = {https://link.aps.org/doi/10.1103/m9w1-ppqz}
}

@article{marcos,
  title={Optimization of entanglement harvesting with arbitrary temporal profiles: the limit of second order perturbation theory},
  author={Morote-Balboa, Marcos and Perche, T Rick},
  journal={arXiv:2604.06303},
  year={2026}
}

@article{summers_bells_1987,
    title = {Bell’s inequalities and quantum field theory. {II}. {Bell}’s inequalities are maximally violated in the vacuum},
    volume = {28},
    issn = {0022-2488},
    url = {https://aip.scitation.org/doi/10.1063/1.527734},
    doi = {10.1063/1.527734},
    number = {10},
    urldate = {2022-07-15},
    journal = {Journal of Mathematical Physics},
    publisher = {American Institute of Physics},
    author = {Summers, Stephen J. and Werner, Reinhard},
    month = oct,
    year = {1987},
    pages = {2448--2456},
}

@article{summers_maximal_1987,
    title = {Maximal violation of {Bell}'s inequalities is generic in quantum field theory},
    volume = {110},
    issn = {0010-3616, 1432-0916},
    url = {https://projecteuclid.org/journals/communications-in-mathematical-physics/volume-110/issue-2/Maximal-violation-of-Bells-inequalities-is-generic-in-quantum-field/cmp/1104159237.full},
    number = {2},
    urldate = {2023-01-04},
    journal = {Communications in Mathematical Physics},
    publisher = {Springer},
    author = {Summers, Stephen J. and Werner, Reinhard},
    month = jan,
    year = {1987},
    pages = {247--259},
}

@article{klco_entanglement_2022,
  title = {Entanglement structures in quantum field theories: Negativity cores and bound entanglement in the vacuum},
  author = {Klco, Natalie and Beck, D. H. and Savage, Martin J.},
  journal = {Phys. Rev. A},
  volume = {107},
  issue = {1},
  pages = {012415},
  numpages = {22},
  year = {2023},
  month = {Jan},
  publisher = {American Physical Society},
  doi = {10.1103/PhysRevA.107.012415},
  url = {https://link.aps.org/doi/10.1103/PhysRevA.107.012415}
}

@article{bianchi_entropy_2019,
    title = {Entropy of a subalgebra of observables and the geometric entanglement entropy},
    volume = {99},
    issn = {2470-0010, 2470-0029},
    url = {http://arxiv.org/abs/1901.06454},
    doi = {10.1103/PhysRevD.99.085001},
    number = {8},
    urldate = {2021-06-22},
    journal = {Physical Review D},
    author = {Bianchi, Eugenio and Satz, Alejandro},
    month = apr,
    year = {2019},
    note = {arXiv: 1901.06454},
    pages = {085001},
}

@article{Polo-Gomez:2023gaz,
    author = "Polo-G{\'o}mez, Jos{\'e} and Mart{\'\i}n-Mart{\'\i}nez, Eduardo",
    title = "{Nonperturbative method for particle detectors with continuous interactions}",
    doi = "10.1103/PhysRevD.109.045014",
    journal = "Phys. Rev. D",
    volume = "109",
    number = "4",
    pages = "045014",
    year = "2024"
}

@article{EntEmbezz,
  title = {Relativistic Quantum Fields Are Universal Entanglement Embezzlers},
  author = {van Luijk, Lauritz and Stottmeister, Alexander and Werner, Reinhard F. and Wilming, Henrik},
  journal = {Phys. Rev. Lett.},
  volume = {133},
  issue = {26},
  pages = {261602},
  numpages = {8},
  year = {2024},
  month = {Dec},
  publisher = {American Physical Society},
  doi = {10.1103/PhysRevLett.133.261602},
  url = {https://link.aps.org/doi/10.1103/PhysRevLett.133.261602}
}

@misc{sorkinAreaLawOG,
      title={1983 paper on entanglement entropy: "On the Entropy of the Vacuum outside a Horizon"}, 
      author={Rafael D. Sorkin},
      year={2014},
      eprint={1402.3589},
      archivePrefix={arXiv},
      primaryClass={gr-qc},
      url={https://arxiv.org/abs/1402.3589}, 
}

@misc{marcos_prep,
      title={(in preparation)}, 
      author={Morote-Balboa, Marcos and Ribes-Metidieri, Patricia and Rick Perche, T.},
      year={2026}
}

\onecolumngrid

\appendix

{\begin{center}
    

\end{center}
}
\section{Laplacian eigenfunctions and smeared field correlations\label{app:calculations}}

In this appendix we will explicitly develop the expressions for the integrands of the field and momentum correlations smeared against the basis of eigenfunctions of the Laplacian in two spheres with Dirichlet boundary conditions. We use the following conventions for a massless scalar quantum field $\hat{\phi}(\mf x)$, its evaluation at the Cauchy surface $t=0$, $\hat{\Phi}(\bm x)$, and its conjugate momentum at this surface, $\hat{\Pi}(\bm x) = \partial_t \hat{\phi}(\mf x)|_{t=0}$:

\begin{equation}
    \hat{\phi}(\mf x) = \frac{1}{(2\pi)^\frac{3}{2}}\int \frac{\dd^3\bm k}{\sqrt{2 |\bm k|}} \left(e^{\ii \mf k \cdot \mf x}\hat{a}_{\bm k} + e^{-\ii \mf k \cdot \mf x}\hat{a}_{\bm k}^\dagger\right),
\end{equation}

\begin{equation}
    \hat{\Phi}(\bm x) = \frac{1}{(2\pi)^\frac{3}{2}}\int \frac{\dd^3\bm k}{\sqrt{2 |\bm k|}} \left(e^{\ii \bm k \cdot \bm x}\hat{a}_{\bm k} + e^{-\ii \bm k \cdot \bm x}\hat{a}_{\bm k}^\dagger\right), \quad \quad \quad
    \hat{\Pi}(\bm x) = \ii \frac{1}{(2\pi)^\frac{3}{2}}\int \dd^3\bm k \sqrt{\frac{|\bm k|}{2}} \left(-e^{\ii \bm k \cdot \bm x}\hat{a}_{\bm k} + e^{-\ii \bm k \cdot \bm x}\hat{a}_{\bm k}^\dagger\right).
\end{equation}
Denoting the smeared field and momentum operators as
\begin{equation}
    \hat{\Phi}(F) = \int \dd^3 \bm x\, \hat{\Phi}(\bm x) F(\bm x), \quad \quad \quad \hat{\Pi}(G) = \int \dd^3 \bm x \,\hat{\Pi}(\bm x) G(\bm x),
\end{equation}
we can write the expected value of the field and momentum anti-commutators relevant for the computation of the vacuum covariance matrix as integrals in Fourier space
\begin{equation}\label{eq:anticomm}
    \langle \{\hat{\Phi}(F),\hat{\Phi}(G)\}\rangle = \frac{1}{(2\pi)^3}\int \frac{\dd^3 \bm k}{|\bm k|} \tilde{F}(\bm k) \tilde{G}(-\bm k), \quad \quad \quad
    \langle \{\hat{\Pi}(F),\hat{\Pi}(G)\}\rangle = \frac{1}{(2\pi)^3}\int \dd^3 \bm k\, |\bm k| \tilde{F}(\bm k) \tilde{G}(-\bm k),
\end{equation}
where we use the following convention for the Fourier transform
\begin{equation}
    \tilde{F}(\bm k) \coloneqq \int \dd^3 \bm x \, F(\bm x) e^{\ii \bm k \cdot \bm x}.
\end{equation}

As discussed in the main text, we consider bases of eigenfunctions of the Laplacian in a ball of radius $R$, $B_R\subset \mathbb{R}^3$, with Dirichlet boundary conditions. Using spherical coordinates we can separate the Laplacian, looking for solutions $u(r,\theta,\phi) = R(r)Y(\theta,\phi)$ satisfying
\begin{equation}
    - \frac{1}{R(r)Y(\theta,\phi)}\nabla^2 R(r)Y(\theta,\phi) = - \frac{1}{R}\frac{1}{r^2}\pdv{}{r}\left(r^2\pdv{R}{r}\right) - \frac{1}{r^2} \frac{\text{L}^2 Y(\theta,\phi)}{Y(\theta,\phi)} = \lambda^2,
\end{equation}
where $\text{L}^2$ is the squared angular momentum operator, with eigenvalues $l(l+1)$, and spherical harmonics as eigenfunctions, $Y_{lm}(\theta,\phi)$, with $l = 0,1,...$ and $m = -l,-l+1,...,l-1,l$. Choosing $Y = Y_{lm}$, we find
\begin{equation}
    r^2 R''(r) + 2r R'(r) + (\lambda^2r^2 + l(l+1))R(r) = 0,
\end{equation}
which is a spherical Bessel equation for $R(r)$, with solutions
\begin{equation}
    R(r) = j_l(\lambda r) = \sqrt{\frac{\pi}{2}}\frac{J_{l+1/2}(\lambda r)}{\sqrt{\lambda r}},
\end{equation}
which are regular at the origin. Here $J_n(x)$ denotes the Bessel function of the first kind. The Dirichlet boundary conditions imposed at $r = R$ give us the possible values of $\lambda$, through the condition $J_l(\lambda R) = 0$. Labelling the $n$th zero of the spherical Bessel function $j_l(z)$ by $\alpha_{nl}$, we find the eigenvalues
\begin{equation}
    \lambda_{nl} = \frac{\alpha_{nl}}{R}.
\end{equation}
For convenience, we will use a real basis of eigenfunctions for the Laplacian, obtaining a real basis of $L^2(B_R)$. The real spherical harmonics can be written as
\begin{equation}
    y_{lm}(\theta,\phi) = \begin{cases}
        \frac{(-1)^m}{\sqrt{2}}(Y_{lm}(\theta,\phi) + Y_{lm}^*(\theta,\phi)), &m>0\\
        Y_{lm}(\theta,\phi), &m=0\\
        \frac{(-1)^m}{\ii \sqrt{2}}(Y_{l|m|}(\theta,\phi) - Y_{l|m|}^*(\theta,\phi)), &m>0
    \end{cases}
\end{equation}
Our orthonormal basis will then be given by
\begin{equation}
    F_{nlm}(r,\theta,\phi) = N_{nlm} j_l(\alpha_{nl} r/R) y_{lm}(\theta,\phi),
\end{equation}
where $N_{nlm}$ are positive normalization constants ensuring that $\{F_{nlm}\}$ is an orthonormal basis of $L^2(B_R)$. We can find $N_{nlm}$ explicitly:
\begin{equation}
    ||F_{nlm}||^2 = N_{nml}^2 \int_0^R \dd r \, r^2 j_{l}(\alpha_{nl}r/R)^2 = N_{nml}^2 R^3 \int_0^1 \dd u \, u^2  j_{l}(\alpha_{nl}u)^2.
\end{equation}
This integral can be solved by using~\cite{gradshteyn}
\begin{equation}\label{eq:lommel}
    \int_0^1 \dd u \, u^2 j_l(au) j_l(bu) = \frac{b j_l(a) j_{l-1}(b) - a j_l(b) j_{l-1}(a)}{a^2 - b^2},
\end{equation}
as well as its limit\footnote{The result follows using the identity $j'_l(z) = \frac{l}{z}j_l(z) - j_{l+1}(z)$.} as $b\to a$:
\begin{align}
    \int_0^1 \dd u \, u^2 j_l(au)^2 &= \frac{1}{2}\left(j_l(a)^2 -j_{l-1}(a)j_{l+1}(a)\right),
\end{align}

Imposing $||F_{nlm}||^2 = 1 $, we find
\begin{equation}
    N_{nlm} = \sqrt{\frac{2}{R^3}}\frac{1}{\sqrt{|j_{l+1}(\alpha_{nl})j_{l-1}(\alpha_{nl})|}}.
\end{equation}
The normalized solutions are then
\begin{equation}
    F_{nlm}(r,\theta,\phi) = \sqrt{\frac{2}{R^3}}\frac{j_l(\alpha_{nl}r/R)}{\sqrt{|j_{l+1}(\alpha_{nl})j_{l-1}(\alpha_{nl})|}}y_{lm}(\theta, \phi).
\end{equation}

To express the smeared field and momentum anti-commutators as in Eq.~\eqref{eq:anticomm}, we must also compute the Fourier transform of this basis of functions. Define
\begin{equation}\label{eq:psiFourierTransform}
    \psi_{nlm}(\bm k) = \int \dd^3 \bm x F_{nlm}(\bm x) e^{\ii \bm k \cdot \bm x}.
\end{equation}
The integral above can be computed utilizing the Rayleigh expansion in terms of real spherical harmonics:
\begin{equation}
    e^{\ii \bm k \cdot \bm x} = 4\pi \sum_{l=0}^\infty \sum_{m=-l}^l \ii^l j_l(k r) y_{lm}(\theta,\phi)y_{lm}(\vartheta,\varphi),
\end{equation}
where we denote the spherical coordinates of $\bm k$ by $(k,\vartheta,\varphi)$. Integrating Eq.~\eqref{eq:psiFourierTransform} in the spherical angles, orthogonality of spherical harmonics gives us
\begin{align}
    \psi_{nlm}(k,\vartheta,\varphi) &= \int_0^R \dd r \,r^2 \int_0^\pi \!\! \dd \theta \int_0^{2\pi}\!\!\!\!\dd\phi \sin\theta \sqrt{\frac{2}{R^3}}\frac{j_l(\alpha_{nl}r/R)}{|j_{l+1}(\alpha_{nl})|}y_{lm}(\theta, \phi)\times 4\pi \sum_{l'=0}^\infty \sum_{m'=-l'}^{l'} \ii^{l'}j_{l'}(k r) y_{lm}(\theta,\phi)y_{l'm'}(\vartheta,\varphi)\nonumber\\
    &= \sqrt{\frac{2}{R^3}}\sum_{l'=0}^\infty\sum_{m'=-l'}^{l'}  \frac{4\pi\ii^l}{|j_{l+1}(\alpha_{nl})|}y_{l'm'}(\vartheta,\varphi)\int_0^R \dd r \,r^2 j_l(\alpha_{nl}r/R)j_{l'}(k r) \int_0^\pi \!\! \dd \theta \int_0^{2\pi}\!\!\!\!\dd\phi \sin\theta  y_{lm}(\theta, \phi)  y_{l'm'}(\theta,\phi)\nonumber\\
    &= \sqrt{\frac{2}{R^3}}\frac{4\pi \ii^l }{|j_{l+1}(\alpha_{nl})|}y_{lm}(\vartheta, \varphi) \int_0^R \dd r \,r^2 j_l(\alpha_{nl}r/R)j_l(kr).
\end{align}
the remaining integral in $r$ is solved using~\eqref{eq:lommel}:
\begin{equation}
    \int_0^R \dd r \,r^2 j_l(\alpha_{nl}r/R)j_l(kr) = R^3 \int \dd u\, u^2 j_l(\alpha_{nl}u)j_l(kR u) =-  R^3 \frac{\alpha_{nl} j_l(kR) j_{l-1}(\alpha_{nl})}{(kR)^2 - \alpha_{nl}^2},
\end{equation}
so that the Fourier transform reads
\begin{equation}
    \psi_{nlm}(k,\vartheta,\varphi) = -  4\pi \ii^l \alpha_{nl} \frac{j_{l-1}(\alpha_{nl})}{\sqrt{|j_{l+1}(\alpha_{nl})j_{l-1}(\alpha_{nl})|}} \sqrt{2R^3} \frac{j_l(kR)  }{(kR)^2 - \alpha_{nl}^2} y_{lm}(\vartheta, \varphi).
\end{equation}

To compute $\psi_{nlm}(-\bm k)$, we note that the reflection of a real spherical harmonic is given by
\begin{equation}
    y_{lm}(-\hat{\bm x}) = (-1)^ly_{lm}(\hat{\bm x}) \iff y_{lm}(\pi - \theta,\pi + \phi) = (-1)^l y_{lm}(\theta,\phi),
\end{equation}
so that
\begin{align}
    \psi_{nlm}(-\bm k) &= -  4\pi \ii^l(-1)^l \alpha_{nl} \frac{j_{l-1}(\alpha_{nl})}{\sqrt{|j_{l+1}(\alpha_{nl})j_{l-1}(\alpha_{nl})|}} \sqrt{2R^3} \frac{j_l(kR)  }{(kR)^2 - \alpha_{nl}^2} y_{lm}(\vartheta, \varphi)\\
    &= -  4\pi \ii^{-l} \alpha_{nl} \frac{j_{l-1}(\alpha_{nl})}{\sqrt{|j_{l+1}(\alpha_{nl})j_{l-1}(\alpha_{nl})|}} \sqrt{2R^3} \frac{j_l(kR)  }{(kR)^2 - \alpha_{nl}^2} y_{lm}(\vartheta, \varphi).
\end{align}

We can now recast the expressions for the field and momentum anti-commutators smeared against different functions in this basis. This will give us the corresponding components of the local covariance matrices $\bm \sigma_\tc{aa} = \bm \sigma_\tc{bb}$ in Eq.~\eqref{eq:cov} of the main text. Plugging the expressions for $\psi_{n_1l_1m_1}(\bm k)$ and $\psi_{n_2l_2m_2}(-\bm k)$ into Eq.~\eqref{eq:anticomm} yields 
\begin{align}
    \langle \{\hat{\Phi}(F_{n_1l_1m_1}),\hat{\Phi}(F_{{n_2l_2m_2}})\}\rangle &=\frac{1}{(2\pi)^3}\int \frac{\dd^3 \bm k}{2|\bm k|} \psi_{n_1l_1m_1}(\bm k) \tilde{\psi}_{n_2l_2m_2}(-\bm k) \\
    &=\frac{(4\pi)^2}{(2\pi)^3} R^3 \alpha_{n_1l_1}\alpha_{n_2l_2}\delta_{l_1l_2}\delta_{m_1m_2} \frac{j_{l_1-1}(\alpha_{n_1l_1})}{\sqrt{|j_{l_1+1}(\alpha_{n_1l_1})j_{l_1-1}(\alpha_{n_1l_1})|}}\frac{j_{l_2-1}(\alpha_{n_2l_2})}{\sqrt{|j_{l_2+1}(\alpha_{n_2l_2})j_{l_2-1}(\alpha_{n_2l_2})|}}\nonumber\\
    &\quad\quad \times \int \dd k  \frac{k\,j_{l_1}(kR)j_{l_2}(kR)}{((kR)^2 - (\alpha_{n_1l_1})^2)((kR)^2 - (\alpha_{n_2l_2})^2)}\nonumber \\
    &=\frac{2R^3}{\pi} \alpha_{n_1l_1}\alpha_{n_2l_2}\delta_{l_1l_2}\delta_{m_1m_2} \frac{j_{l_1-1}(\alpha_{n_1l_1})}{\sqrt{|j_{l_1+1}(\alpha_{n_1l_1})j_{l_1-1}(\alpha_{n_1l_1})|}}\frac{j_{l_2-1}(\alpha_{n_2l_2})}{\sqrt{|j_{l_2+1}(\alpha_{n_2l_2})j_{l_2-1}(\alpha_{n_2l_2})|}}\nonumber\\
    &\quad\quad \times \int \dd k  \frac{k\,j_{l_1}(kR)j_{l_2}(kR)}{((kR)^2 - (\alpha_{n_1l_1})^2)((kR)^2 - (\alpha_{n_2l_2})^2)}.\nonumber
\end{align}
Similarly, for the smeared momentum operator, it is enough to add a factor of $k^2$ to the integral above:
\begin{align}
    \langle \{\hat{\Pi}(F_{n_1l_1m_1}),\hat{\Pi}(F_{{n_2l_2m_2}})\}\rangle &=\frac{2R^3}{\pi} \alpha_{n_1l_1}\alpha_{n_2l_2}\delta_{l_1l_2}\delta_{m_1m_2} \frac{j_{l_1-1}(\alpha_{n_1l_1})}{\sqrt{|j_{l_1+1}(\alpha_{n_1l_1})j_{l_1-1}(\alpha_{n_1l_1})|}}\frac{j_{l_2-1}(\alpha_{n_2l_2})}{\sqrt{|j_{l_2+1}(\alpha_{n_2l_2})j_{l_2-1}(\alpha_{n_2l_2})|}}\\
    &\quad\quad \times \int \dd k  \frac{k^3\,j_{l_1}(kR)j_{l_2}(kR)}{((kR)^2 - (\alpha_{n_1l_1})^2)((kR)^2 - (\alpha_{n_2l_2})^2)}.
\end{align}

When considering modes in two spheres separated by a distance $\bm L$, we consider the shifted basis functions \mbox{$F_{nml}(\bm x) \mapsto F_{nml}(\bm x - \bm L) \equiv F_{nml}^{(\bm L)}(\bm x)$}, resulting in the shifted Fourier transform 
\begin{equation}
    \psi^{(\bm L)}_{nlm}(\bm k) = \int \dd^3 \bm x F_{nlm}(\bm x - \bm L)e^{\ii \bm k \cdot \bm x} = e^{\ii \bm k \cdot \bm L} \psi_{nlm}(\bm k).
\end{equation}
Thus, we obtain the following field and momentum smeared anti-commutators between basis of functions in shifted spheres
\begin{align}
    &\langle \{\hat{\Phi}(F_{n_1l_1m_1}),\hat{\Phi}(F^{(\bm L)}_{{n_2l_2m_2}})\}\rangle =\frac{1}{(2\pi)^3}\int \frac{\dd^3 \bm k}{2|\bm k|} \psi_{n_1l_1m_1}(\bm k)e^{\ii \bm k \cdot \bm L} {\psi}_{n_2l_2m_2}(-\bm k) \\
    &=\frac{(4\pi)^2}{(2\pi)^3} R^3 \alpha_{n_1l_1}\alpha_{n_2l_2} \frac{j_{l_1-1}(\alpha_{n_1l_1})}{\sqrt{|j_{l_1+1}(\alpha_{n_1l_1})j_{l_1-1}(\alpha_{n_1l_1})|}}\frac{j_{l_2-1}(\alpha_{n_2l_2})}{\sqrt{|j_{l_2+1}(\alpha_{n_2l_2})j_{l_2-1}(\alpha_{n_2l_2})|}}\ii^{l_1-l_2}\nonumber\\
    &\quad\quad\quad\quad\quad\quad\quad\quad\quad\times \int \dd k  \frac{k\,j_{l_1}(kR)j_{l_2}(kR)}{((kR)^2 - (\alpha_{n_1l_1})^2)((kR)^2 - (\alpha_{n_2l_2})^2)} \int \dd \Omega y_{l_1m_1}(\vartheta,\varphi)y_{l_2m_2}(\vartheta,\varphi)e^{\ii \bm k \cdot \bm L}\nonumber,\\
    &\langle \{\hat{\Pi}(F_{n_1l_1m_1}),\hat{\Pi}(F^{(\bm L)}_{{n_2l_2m_2}})\}\rangle =\frac{1}{(2\pi)^3}\int \frac{\dd^3 \bm k}{2|\bm k|} \psi_{n_1l_1m_1}(\bm k)e^{\ii \bm k \cdot \bm L} {\psi}_{n_2l_2m_2}(-\bm k) \\
    &=\frac{(4\pi)^2}{(2\pi)^3} R^3 \alpha_{n_1l_1}\alpha_{n_2l_2} \frac{j_{l_1-1}(\alpha_{n_1l_1})}{\sqrt{|j_{l_1+1}(\alpha_{n_1l_1})j_{l_1-1}(\alpha_{n_1l_1})|}}\frac{j_{l_2-1}(\alpha_{n_2l_2})}{\sqrt{|j_{l_2+1}(\alpha_{n_2l_2})j_{l_2-1}(\alpha_{n_2l_2})|}}\ii^{l_1-l_2}\\
    &\quad\quad\quad\quad\quad\quad\quad\quad\quad\times \int \dd k  \frac{k^3\,j_{l_1}(kR)j_{l_2}(kR)}{((kR)^2 - (\alpha_{n_1l_1})^2)((kR)^2 - (\alpha_{n_2l_2})^2)} \int \dd \Omega y_{l_1m_1}(\vartheta,\varphi)y_{l_2m_2}(\vartheta,\varphi)e^{\ii \bm k \cdot \bm L}\nonumber.
\end{align}
The results above allow on to compute components of the cross-correlation matrix $\bm \sigma_\tc{ab}$ in Eq.~\eqref{eq:cov} of the main text, assuming that sphere B is separated from sphere A by a distance of $\bm L$, while $\bm \sigma_\tc{ba} = \bm \sigma_{\tc{ab}}^{\bm \top}$.

For concreteness, we end the appendix by stating how the expressions above simplify when we restrict ourselves to the case $m = 0$, assuming that the $z$ axis is aligned with the separation vector $\bm L$. In this case, we define the function
\begin{align}
    \mathcal{I}_{l_1,l_2}(\bm k\cdot \bm L) \coloneqq \int \dd \Omega y_{l_10}(\vartheta,\varphi)y_{l_20}(\vartheta,\varphi)e^{\ii \bm k \cdot \bm L} = 2\pi \int_0^\pi \dd \theta y_{l_10}(\vartheta,0)y_{l_20}(\vartheta,0)e^{\ii k |\bm L| \cos\theta} \equiv \mathcal{I}_{l_1,l_2}(k |\bm L|).
\end{align}
The terms $\mathcal{I}_{l1,l2}(k |\bm L|)$ can be written in terms of Wigner's 3j-symbols, but the expression is not particularly insightful, so we omit it\footnote{In the numerical computations we pre-compute the functions $I_{l_1,l_2}$ for the corresponding values of $l$ considered.}. We can then write the smeared anti-commutators as
\begin{align}
    \langle \{\hat{\Phi}(F_{n_1l_10}),\hat{\Phi}(F^{(\bm L)}_{{n_2l_20}})\}\rangle &= 4 R^3 \alpha_{n_1l_1}\alpha_{n_2l_2} \frac{j_{l_1-1}(\alpha_{n_1l_1})}{\sqrt{|j_{l_1+1}(\alpha_{n_1l_1})j_{l_1-1}(\alpha_{n_1l_1})|}}\frac{j_{l_2-1}(\alpha_{n_2l_2})}{\sqrt{|j_{l_2+1}(\alpha_{n_2l_2})j_{l_2-1}(\alpha_{n_2l_2})|}}\ii^{l_1-l_2}\\
    &\quad\quad\quad\quad\quad\quad\quad\quad\quad\times \int \dd k  \frac{k\,j_{l_1}(kR)j_{l_2}(kR)\mathcal{I}_{l_1,l_2}(k |\bm L|)}{((kR)^2 - (\alpha_{n_1l_1})^2)((kR)^2 - (\alpha_{n_2l_2})^2)}.
\end{align}
\begin{align}
    \langle \{\hat{\Pi}(F_{n_1l_10}),\hat{\Pi}(F^{(\bm L)}_{{n_2l_20}})\}\rangle &= 4 R^3 \alpha_{n_1l_1}\alpha_{n_2l_2} \frac{j_{l_1-1}(\alpha_{n_1l_1})}{\sqrt{|j_{l_1+1}(\alpha_{n_1l_1})j_{l_1-1}(\alpha_{n_1l_1})|}}\frac{j_{l_2-1}(\alpha_{n_2l_2})}{\sqrt{|j_{l_2+1}(\alpha_{n_2l_2})j_{l_2-1}(\alpha_{n_2l_2})|}}\ii^{l_1-l_2}\\
    &\quad\quad\quad\quad\quad\quad\quad\quad\quad\times \int \dd k  \frac{k^3\,j_{l_1}(kR)j_{l_2}(kR)\mathcal{I}_{l_1,l_2}(k |\bm L|)}{((kR)^2 - (\alpha_{n_1l_1})^2)((kR)^2 - (\alpha_{n_2l_2})^2)}.
\end{align}

\section{Regularity and Convergence of the MEMs and their Entanglement}\label{app:convergence_mems}

In this appendix, we investigate the regularity and the $L^2$ convergence of the approximate Most Entangled Modes (MEMs) as the dimension of the subspaces used in their construction is increased. We argue that, while the momentum smearings $G^{\bd}_{N}$ display convergent asymptotics, the position smearings $F^{\bd}_N$ do not converge to an $L^2$ function in the limit of infinite modes $N\to \infty$. This corresponds to the expansion of an integrable function that exhibits non-square integrable boundary divergences, in analogy with the analytical results obtained in 1+1 dimensions~\cite{Jason}. Importantly, even if the field smearings $F^{\bd}_N$ do not converge in $L^2$, the physical quantities derived from it---such as the LN---remain well-defined in the limit $N \to \infty$ due to the convergence of the associated smeared field anti-commutator two-point functions.


\subsection{Numerical convergence of the MEM profiles}

We obtained numerical approximations to the MEMs of two disjoint spherical regions by applying the entanglement partner construction described in~\cite{klco_entanglement_2022,partnerformula} to subspaces of increasing dimensions. The underlying modes  are obtained from a basis $\{\psi_{nl0}\}$ of $L^2$ (axisymmetric) functions satisfying Dirichlet boundary conditions on each spherical region, truncated to a finite number of modes $n= 1, \dots, N$ and $l=0, \dots, L$. By changing $N$ and $L$ we obtain a sequence of finite-dimensional approximations to the axisymmetric MEMs, $F^{\bd}_{NL}$ and $G^{\bd}_{NL}$, defined by 
\begin{equation}
    F^{\bdiamond}_{NL} =  \sum_{n=1}^N \sum_{l=0}^Lq_{nl}^{NL} \psi_{nl0}\,, \quad   G^{\bdiamond}_{NL} =  \sum_{n=1}^N \sum_{l=0}^Lp_{nl}^{NL} \psi_{nl0}\,,
\end{equation}
where the coefficients $q^{NL}_{nl}$ and $p_{nl}^{NL}$ are obtained by solving the eigenvalue problem for the partial transposed complex structure $J^\Gamma$ within the subspace spanned by $\{\psi_{nl0}\}_{n\in\{1,..., N\}, l\in\{0,...,L\}}$.



Figure~\ref{fig:GMEM_different_N} shows the momentum smearing $G^{\bdiamond}_{NL}$ in region $\Sigma_\tc{a}$ for $L=L_{\text{max}}\equiv 30$ and different values of $N$, for a fixed separation $d =  R/2$ between the spherical regions $\Sigma_{\tc{a}}$ and $\Sigma_\tc{b}$. We observe that as $N$ increases, the differences between the profiles  $G^{\bdiamond}_{NL_{\text{max}}}$ become progressively smaller, approaching a stable shape. This suggests that the momentum smearing approaches a well-defined limiting profile as the dimension of the truncated subspace increases. Indeed, in the following we will argue that the corresponding coefficients in the numerical expansion for $G^{\bd}_{NL}$ converge in $L^2$. 

\begin{figure*}[h!]
    \centering
    \begin{tikzpicture}
       \node[anchor=west] (n310) at ([xshift=2cm]current page.west) {\includegraphics[width=0.33\textwidth,trim={1cm 0 1cm 1.5cm}, clip]{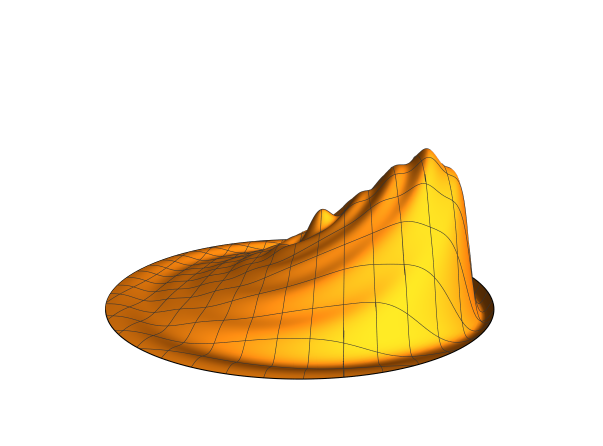}};
       \node[anchor=north] (l) at (n310.south) {$N = 10$}; 
       \node[anchor=center] (n620) at (current page.center) {\includegraphics[width=0.34\textwidth,trim={1cm 0.75cm 1cm 1.75cm}, clip]{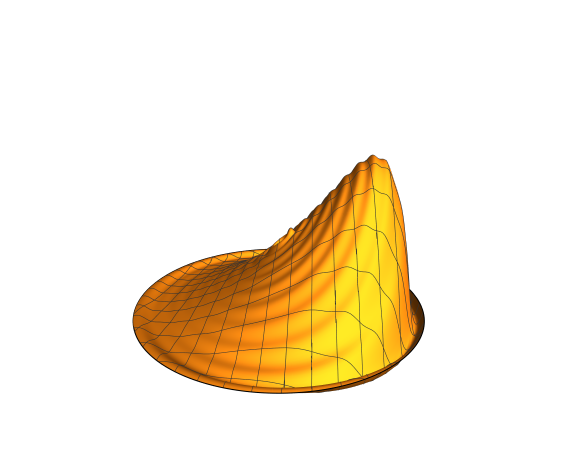}};
       \node[anchor=north] (l2) at (n620.south) {$N = 20$}; 
       \node[anchor=east] (n930)at ([xshift=-2cm]current page.east) {\includegraphics[width=0.33\textwidth,trim={1cm 0.75cm 1cm 1.5cm}, clip]{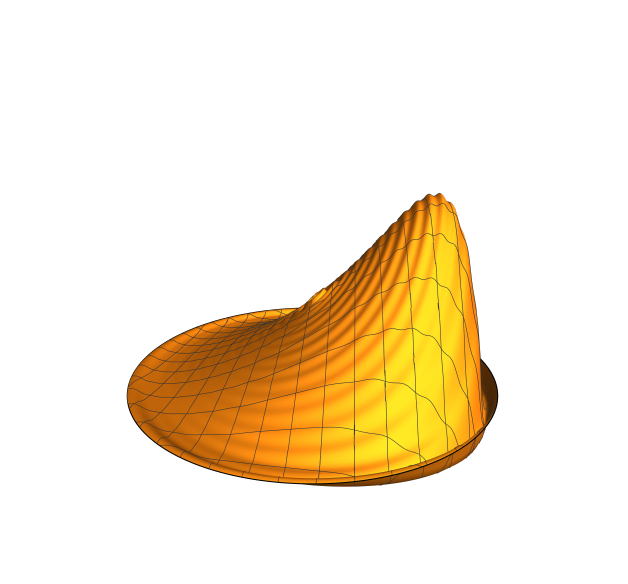}};
       \node[anchor=center] at ([xshift=0.3\textwidth]l2.center) {$N = 30$}; 
    \end{tikzpicture}
    \caption{Profile of momentum smearing for the MEM profile $-G^{\bd}_{NL_{\mathrm{max}}}$ in region $\Sigma_{\tc{a}}$ computed with $N=10, 20, 30$ and $L_{\text{max}} = 30$, modes when the distance between the regions $\Sigma_{\tc{a}}$ and $\Sigma_{\tc{b}}$ is $d=R/2$.  }
    \label{fig:GMEM_different_N}
\end{figure*}
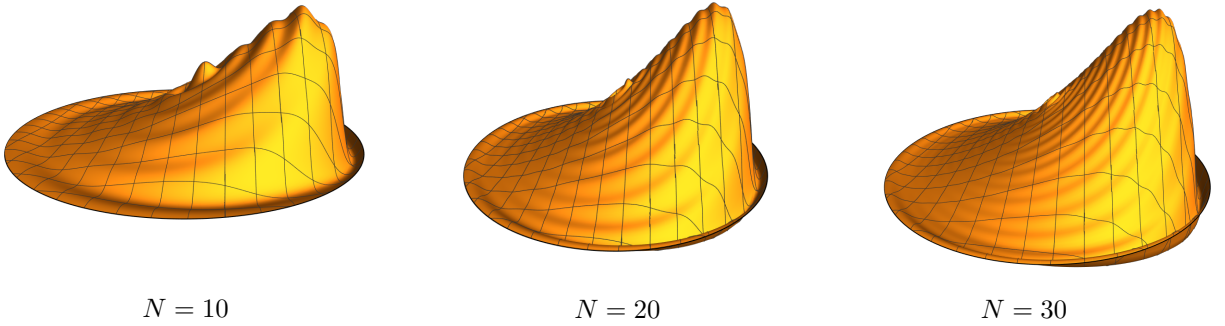

In contrast, the position MEM profiles in region $\Sigma_\tc{a}$, $F^{\bd}_{NL}$ exhibits a qualitatively different behavior. Figure~\ref{fig:FMEM_different_N} shows the profile of the field smearing $F^{\bdiamond}_{NL_{\text{max}}}$ in region $\Sigma_{\tc{a}}$ for increasing $N$, at a fixed separation $d =  R/2$ between the spherical regions. For each fixed $N$, we see that the profile $F^{\bd}_{NL_\text{max}}$ displays $N$ oscillations, consistent with the expansion of an $L^1$ function that is not square integrable. Also notice that as $N$ increases, the boundary oscillations increase, also consistent with the behaviour of a function exhibiting non-square integrable boundary divergences. Overall, the sequence seems to develop structure on progressively smaller spatial scales, suggesting that the sequence $F^{\bdiamond}_{NL}$ does not approach a limiting  $L^2$ profile.  

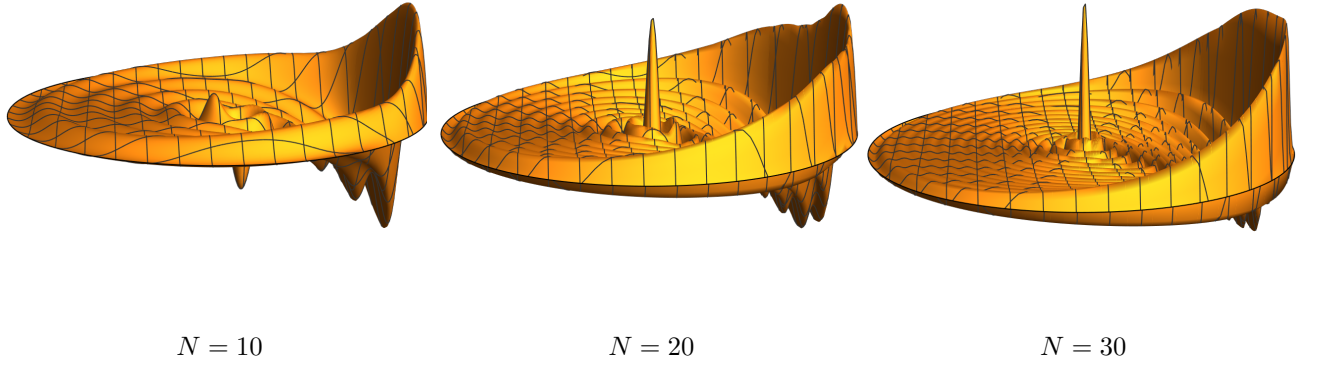
\begin{figure*}[h!]
    \centering
    \begin{tikzpicture}
       \node[anchor=west] (n310) at ([xshift=2cm]current page.west) {\includegraphics[width=0.33\textwidth,trim={0 2cm 0 2cm},clip]{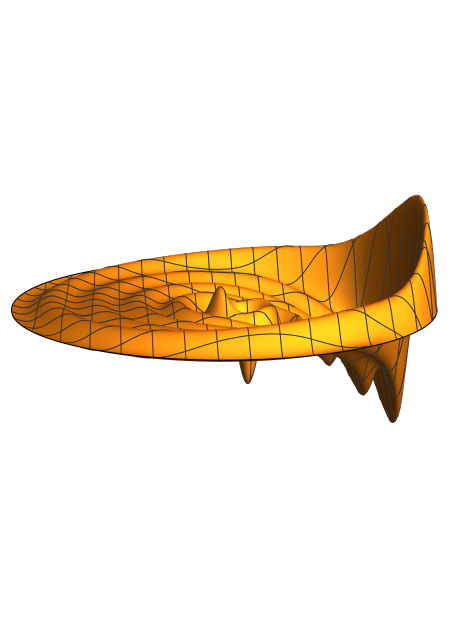}};
       \node[anchor=north] at (n310.south) {$N = 10$}; 
       \node[anchor=center] (n620) at (current page.center) {\includegraphics[width=0.33\textwidth,trim={0 2cm 0 2cm},clip]{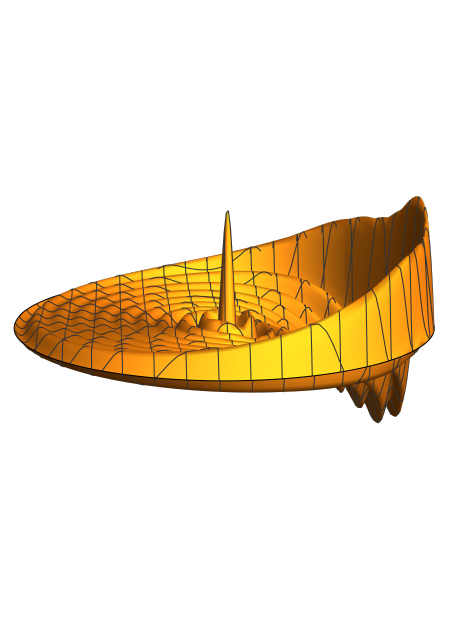}};
       \node[anchor=north] at (n620.south) {$N = 20$}; 
       \node[anchor=east] (n930)at ([xshift=-2cm]current page.east) {\includegraphics[width=0.33\textwidth,trim={0 2cm 0 2cm},clip]{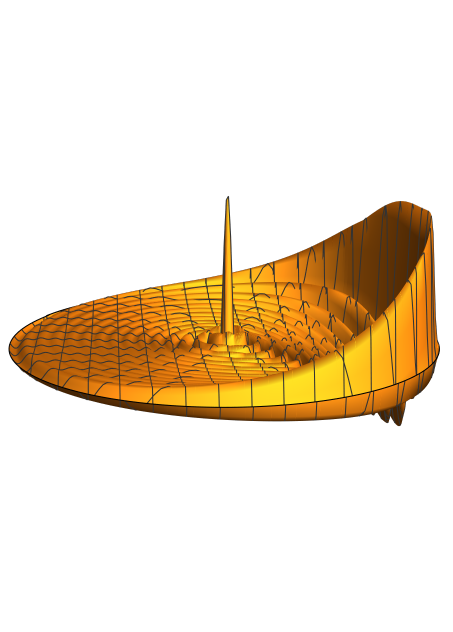}};
       \node[anchor=north] at (n930.south) {$N = 30$}; 
    \end{tikzpicture}
    \caption{ Profile of momentum smearing for the MEM $F^{\bd}_{NL_{\mathrm{max}}}$ in region $\Sigma_{\tc{a}}$ computed with $N=10, 20, 30$ with $L_\text{max} = 30$, modes when the distance between the regions $\Sigma_{\tc{a}}$ and $\Sigma_{\tc{b}}$ is $d=R/2$.  }
    \label{fig:FMEM_different_N}
\end{figure*}

To analyze the convergence of the approximate MEM profiles, we distinguish between the cutoffs for the Laplacian eigenvalues $N$ and $L$. First we argue that $L_{\mathrm{max}} = 30$ is sufficient to capture the angular behavior of the MEMs for the separation values $d$ considered in this work. Next, we argue that while $||G^{\bd}_{NL_{\mathrm{max}}}||_{L^2}$ will remain finite as $N \to \infty$, $||F^{\bd}_{NL_{\mathrm{max}}}||_{L^2} \to \infty$.  

Since the functions $\{\psi_{nl0}\}$ are orthogonal in the $L^2$-norm, the norm of the truncated MEMs can be written as 
$$ ||F^{\bd}_{NL}||_{L^2} = \left( \sum_{n=1}^N \sum_{l=0}^{L} |q_{nl}^{NL}|^2 \right)^{1/2}, \quad  ||G^{\bd}_{NL}||_{L^2} = \left( \sum_{n=1}^N \sum_{l=0}^{L} |p_{nl}^{NL}|^2 \right)^{1/2}.$$
The left panel of Fig.~\ref{fig:qvsl} shows the sum $\sum_{n=1}^N |q_{nl}^{NL}|^2$ when the separation between the regions $\Sigma_\tc{a}$ and $\Sigma_\tc{b}$ is $d=0.1R$ as a function of $l\in\{0,1,...,L_\text{max}\}$, for different truncations of the radial basis elements $N$. We observe that the sum of the coefficients $q_{nl}^L$ decreases exponentially with increasing $l$, where only coefficients with $l\lesssim 15$ are relevant. The right panel of Fig.~\ref{fig:qvsl} shows the sum $\sum_{n=1}^{N_{\text{max}}}|q_{nl}^{\mathrm{max}}|^2$ obtained for the truncation $N_{\mathrm{max}} = 40$ as a function of $l$ for different separations between the spherical regions $\Sigma_{\tc{A}}$ and $\Sigma_{\tc{B}}$. Also in this case, we observe that the contribution to $\sum_{n=1}^{N_{\max}} |q_{nl}^{N_{\mathrm{max}}L}|^2$ of $l\gtrsim 10$ is negligible.
\begin{figure}[h!]
    \centering
    \begin{tikzpicture}
        \node[xshift=-0.25\textwidth] (ql) at (0,0) {\includegraphics[width=0.5\textwidth]{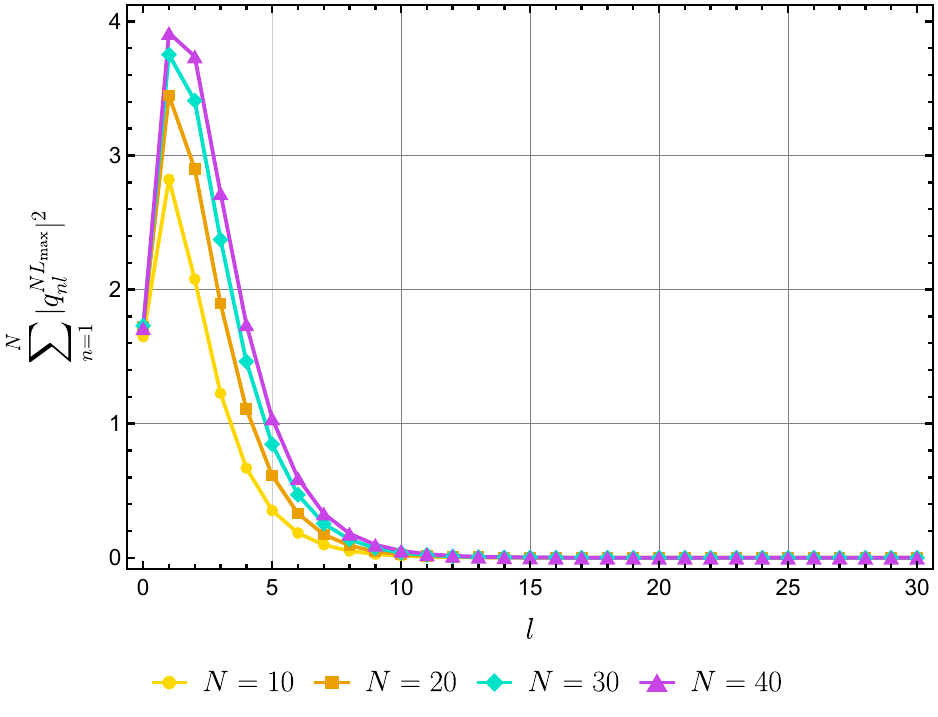}}; 
         \node[xshift=0.25\textwidth] (qr) at (0,0) {\includegraphics[width=0.5\textwidth]{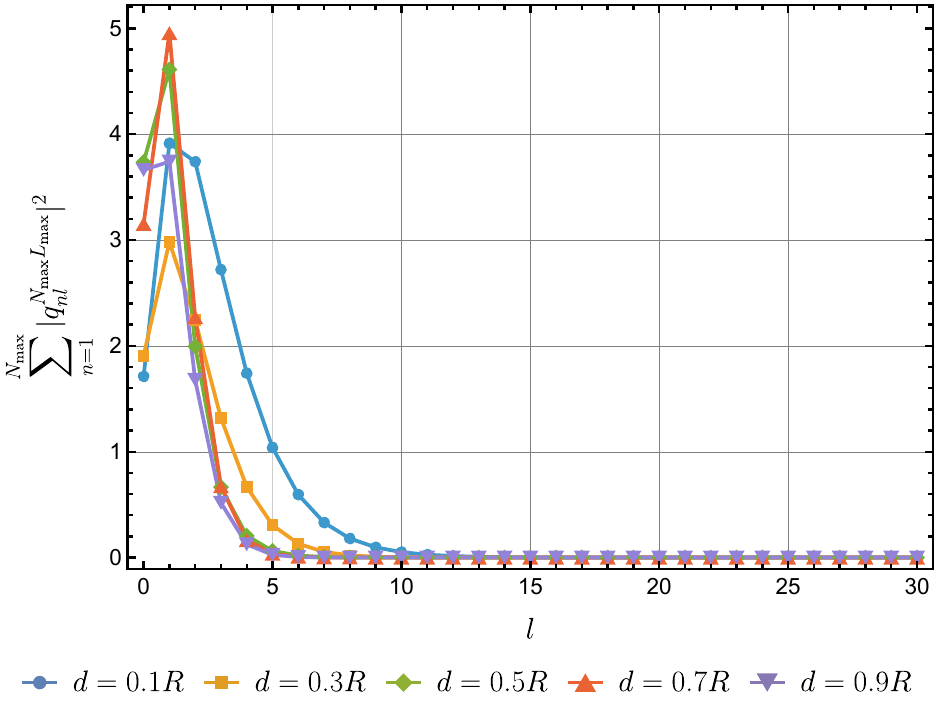}}; 
    \end{tikzpicture}
    \caption{Angular dependence of the coefficients of the field amplitude MEM profile. (Left) The quantity $\sum_{n=1}^N|q_{nl}^{NL_{\mathrm{max}}}|^2$ as a function of $l$ for $d=0.1R$, $L_{\mathrm{max}} = 30$, and for different number of ``radial'' modes $N$. (Right) The quantity $\sum_{n=1}^{N_{\mathrm{max}}} |q_{nl}^{N_{\mathrm{max}} L_{\max}}|^2$ as a function of $l$ for $N_{\max} =40$, $L_{\mathrm{max}} = 30$, for different separations between $\Sigma_{\tc{a}}$ and $\Sigma_{\tc{b}}$.    }
    \label{fig:qvsl}
\end{figure}

Similarly, the left panel of Fig.~\ref{fig:pvsl} shows the sum $\sum_{n=1}^N |p_{nl}^{NL}|^2$ as a function of $l$ when $d=0.1R$ for different truncations and the right panel shows $\sum_{n=1}^{N_{\max}} |p_{nl}^{N_{\max}L}|^2$ as a function of $l$ for $N_{\max} = 40$ and different separations $d$ between the spherical regions. In both cases, we observe that only coefficients with $l\lesssim 15$ meaningfully contribute to $\sum_{n=1}^{N_{\max}} |p_{nl}^{N_{\mathrm{max}}L}|^2$. From Figs.~\ref{fig:qvsl} and~\ref{fig:pvsl}, we conclude that fixing the number of spherical harmonics $l = 0, \dots, L_{\mathrm{max}}$  ($L_{\mathrm{max}} = 30$) accurately captures the behaviour of the MEMs.

\begin{figure}[h!]
    \centering
    \begin{tikzpicture}
        \node[xshift=-0.25\textwidth] (ql) at (0,0) {\includegraphics[width=0.5\textwidth]{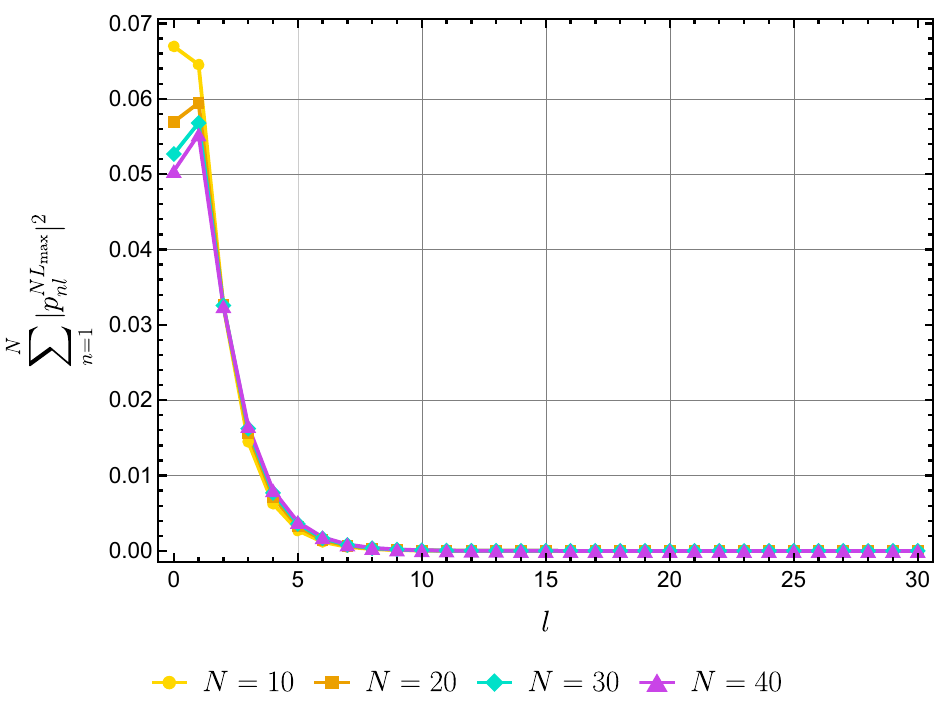}}; 
         \node[xshift=0.25\textwidth] (qr) at (0,0) {\includegraphics[width=0.5\textwidth]{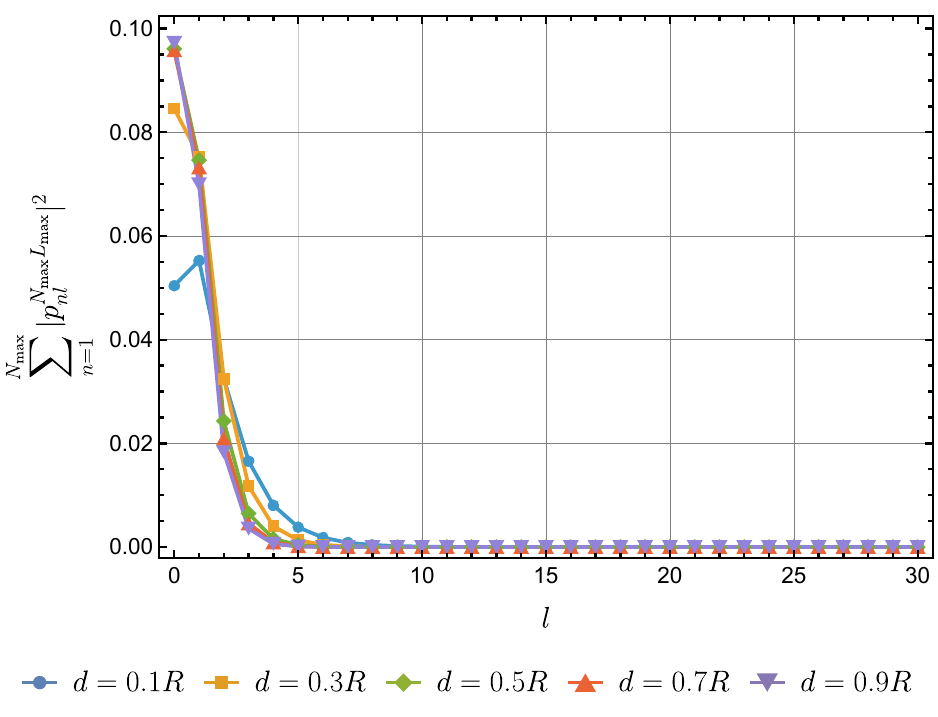}}; 
    \end{tikzpicture}
    \caption{
    {Angular dependence of the coefficients of the momentum MEM profile. (Left) The quantity $\sum_{n=1}^N|p_{nl}^{NL_{\mathrm{max}}}|^2$ as a function of $l$ for $d=0.1R$, $L_{\mathrm{max}} = 30$, and for different number of ``radial'' modes $N$. (Right) The quantity $\sum_{n=1}^{N_{\mathrm{max}}} |p_{nl}^{N_{\mathrm{max}} L_{\max}}|^2$ as a function of $l$ for $N_{\max} =40$, $L_{\mathrm{max}} = 30$, for different separations between $\Sigma_{\tc{a}}$ and $\Sigma_{\tc{b}}$.  } }
    \label{fig:pvsl}
\end{figure}


Next, we study the asymptotic behavior of $||F^{\bd}_{NL_{\mathrm{max}}}||_{L^2}$ and $||G^{\bd}_{N L_{\mathrm{max}}}||_{L^2}$ as $N$ increases. Assume that $H\in L^1(\mathbb{R}^3)$ and that the integrals
\begin{equation}
    h_{nl} := \int \dd^3\bm x \, \psi_{nl0}(\bm x) H(\bm x) 
\end{equation}
are finite for all $n$ and $l$. We also assume that $\sum_{l=0}^\infty |h_{nl}|^2$ converge for all $n$, so that the functions
\begin{equation}
    h_n = \sum_{l=0}^{\infty} h_{nl} \psi_{nl0}
\end{equation}
are square integrable for all $n\geq 1$. We define the sequence of functions
\begin{equation}
    H_N \coloneqq  \sum_{n=1}^{N} h_n.
\end{equation}
Notice that $H_N\to H$ if and only if the sum $\sum_{n=1}^\infty |h_n|^2$ converges, in which case $H\in L^2(\mathbb{R}^3)$. In this sense, $||h_N||^2_{L^2} = ||H_N - H_{N-1}||^2_{L^2}$ can be thought of as the $N$th coefficient of the expansion of $H$ in terms of radial basis functions. In analogy to this expansion, we use the differences $||F^{\bd}_{NL_{\max}} -F^{\bd}_{(N-1)L_{\max}}||_{L^2}$ and $||G^{\bd}_{NL_{\max}} -G^{\bd}_{(N-1)L_{\max}}||_{L^2}$ as a function of $N$ to analyze the asymptotic behavior of the $L^2$-norm of the radial MEMs expansion.

\begin{figure}[h!]
    \centering 
    \begin{tikzpicture}
        \node (fig) at (0,0) {\includegraphics[width=0.75\textwidth,trim={1cm 1cm 0 0}]{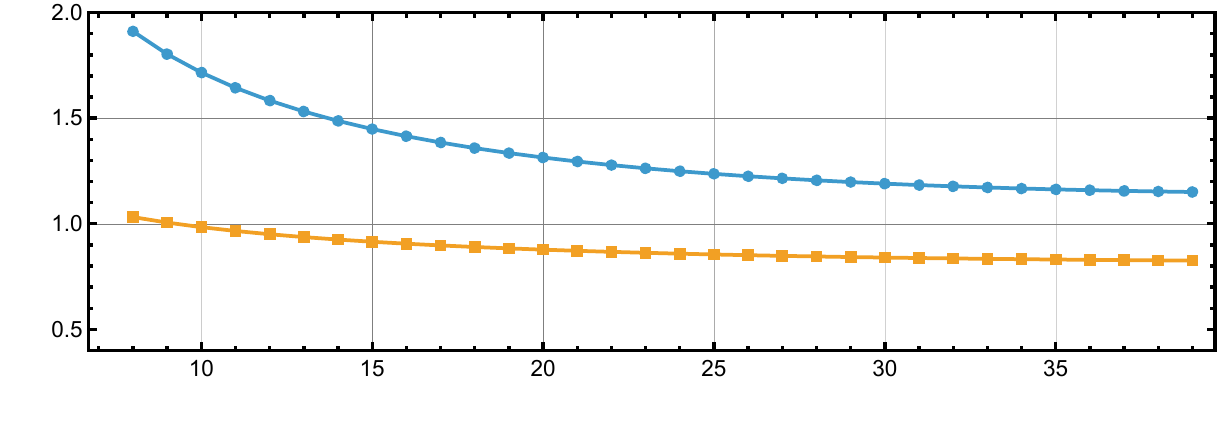}}; 
        \node[anchor=north west] at ([xshift=0cm,yshift=0.625cm]fig.south east) {$N$}; 
        \node[anchor=west] (l)at (fig.east) {\includegraphics[height=1.5cm]{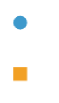}}; 
        \node[anchor=west] at ([yshift=0.5cm,xshift=-0.5cm]l.east) {$\frac{||F^{\bd}_{NL_{\mathrm{max}}}\!\!-F^{\bd}_{(N-1)L_{\mathrm{max}}}||_{L^2}^{-2}}{C_Q^2N}$};
         \node[anchor=west] at ([yshift=-0.5cm,xshift=-0.5cm]l.east) {$\frac{||G^{\bd}_{NL_{\mathrm{max}}}\!\!-G^{\bd}_{(N-1)L_{\mathrm{max}}}||_{L^2}^{-\frac{2}{3}}}{C_P^{2/3}N}$};
\end{tikzpicture}
    \caption{The quantities $(||F^{\bd}_{NL_\text{max}} - F^{\bd}_{(N-1)L_\text{max}}||_{L^2}/C_Q)^{-2} /N$ and $(||G^{\bd}_{NL_\text{max}} - G^{\bd}_{(N-1)L_\text{max}}||_{L^2}/C_P)^{-2/3}/N$ as a function of $N$. 
    }
    \label{fig:asymptotic_behavior}
\end{figure}


 Figure~\ref{fig:asymptotic_behavior} shows $(||F^{\bd}_{NL_\text{max}} - F^{\bd}_{(N-1)L_\text{max}}||_{L^2}/C_Q)^{-2} /N $ and $(||G^{\bd}_{NL_\text{max}} - G^{\bd}_{(N-1)L_\text{max}}||_{L^2}/C_P)^{-2/3} /N$ for constants $C_Q$ and $C_P$ as a function of $N$ for the separation $d=0.1R$. We observe that these two quantities become approximately constant for large $N$. This behavior is consistent with the asymptotic scalings 
\begin{align}
    ||F^{\bd}_{NL_\text{max}} - F^{\bd}_{(N-1)L_\text{max}}||_{L^2} &\sim C_Q/N^{1/2},\\
    ||G^{\bd}_{NL_\text{max}} - G^{\bd}_{(N-1)L_\text{max}}||_{L^2} &\sim C_P/N^{3/2}
\end{align}
 for large $N$.

The scalings for the field and momentum smearings have qualitatively different implications regarding their $L^2$ convergence. For $G^{\bdiamond}_{NL_{\mathrm{max}}}$, the corresponding radial coefficients scale as $N^{-3/2}$, which is square-summable, suggesting  that the $L^2$ norm of the MEMs will be finite as $N\to\infty$. In contrast, for $F^{\bdiamond}_{NL_\text{max}}$, the corresponding coefficients scale as $N^{-1/2}$, giving rise to a sequence that is not square-summable, implying that the limit of  $F^{\bd}_{NL_\text{max}}$ as $N\to \infty$ does not converge to a square integrable function. As we will discuss in the following subsection, this is consistent with the analytical results for MEMs of massless fields in 1+1 dimensional spacetimes found in~\cite{Jason}. 


\subsubsection{Insights from the $1+1$-dimensional case}

While the exact MEMs for two spherical regions in $1+3$ dimensions are not known,  closed-form expressions for the exact MEMs between two intervals of a massless scalar field in $1+1$ dimensions have been found in~\cite{Jason}.  In this subsection, we briefly review their results and discuss the fact that the field amplitude smearing of MEMs in 1+1 dimensions are not square integrable. We also discuss the analogy between this case and the $1+3$ dimensional results found in this work. 

To avoid the well-known infrared divergence associated with the zero mode of a massless scalar field in $1+1$ dimensions, the analysis of~\cite{Jason} restricts the allowed field smearings to functions with zero integral over the real line. The right- and left-moving sectors then decouple, and the corresponding MEMs can be obtained in closed form.

To ease the calculations and exploit the symmetry between left- and right- moving sectors, the field is written as $\hat{\phi}(\mf x) = \hat{\phi}_R(\mf x) + \hat{\phi}_L(\mf x)$. The corresponding conjugate momentum is given by $\hat{\pi}(\mf x) = - \hat{\phi}_R'(\mf x)$ + $\hat{\phi}_L'(\mf x)$, where primes denote derivative with respect to the spatial coordinate $x$ and $\mf x = (t,x)$ in 1+1 dimensions. To make the connection with the notation in this work explicit, we notice that, at a given surface of constant $t$, the field and momentum operator-valued distributions can be written in terms of the right- and left-moving sectors as 
\begin{align}
    \hat{\Phi}(F) &= \hat{\Phi}_R(F) + \hat{\Phi}_L(F),\\
    \hat{\Pi}(G) &= - \hat{\Phi}_R'(G) + \hat{\Phi}'_L(G) = \hat{\Phi}_R(G') -
\hat{\Phi}_L(G'),
\end{align}
where integration by parts in $x$ has been used to write the last expression for the smeared momentum operator. 

The analysis of~\cite{Jason} yields the exact MEM between two intervals $A = [-R-d,-d]$ and $B = [d, d+R]$ of size $R$ separated by a distance $2d$ within the left/right decomposition. In particular, they find the following expressions for the MEM smearings   
\begin{align}
     \memf_{\tc{a},c}(x) &= \frac{C\cos(\alpha \log((x+d+R)(x-d-R))/(x-d)(x+d)))}{\sqrt{-(x+d)(x+d+R)(x-d)(x-d-R)}},\\
    \memf_{\tc{a},s}(x) &= -\frac{C\sin(\alpha \log((x+d+R)(x-d-R))/(x-d)(x+d)))}{\sqrt{-(x+d)(x+d+R)(x-d)(x-d-R)}},
\end{align}
and similarly for the MEMs in $B$, where $C$ is a constant and $\alpha$ is the smallest positive solution to  
\begin{equation}
   P_{-\ii \alpha - \frac{1}{2}}((1+\eta)/(1-\eta)) = 0,\quad \quad \text{with}\quad\quad  \eta = \frac{1}{\left(1 + \frac{2d}{R}\right)^2}.
\end{equation}
The corresponding MEMs are $(\hat{\Phi}_R(\memf_{\tc{a},c}),\hat{\Phi}_R(\memf_{\tc{a},s}))$, $(\hat{\Phi}_R(\memf_{\tc{b},c}),\hat{\Phi}_R(\memf_{\tc{b},s}))$, as well as their identical left moving counterparts. 

Near any boundary of the interval, writing $h$ for the distance to that boundary, these modes behave as
\begin{equation}
F^{\bd}_{A,s}(h)\sim h^{-1/2}\sin\left(\alpha\log(\beta h)\right), \quad F^{\bd}_{A,c}(h)\sim h^{-1/2}\cos\left(\alpha\log (\beta h)\right).
\end{equation}
Consequently, $||F^{\bd}_{A,s}||_{L^2}$ and   $||F^{\bd}_{A,c}||_{L^2}$ diverge, implying that the MEM smearings are not elements of $L^2$. In contrast, the indefinite integral of the field amplitude MEM profiles in 1+1 behave as $h^{1/2}\sin(\alpha \log(\gamma h))$ close to the boundaries, yielding square integrable MEM momentum profiles. The combination of the momentum and field MEM profiles yields finite smeared field and momentum correlations, and implies that the MEMs have finite logarithmic negativity~\cite{Jason}. 

The $1+1$ dimensional result provides an explicit example in which the continuum field MEM profiles are not an $L^2$ function, while the vacuum field and momentum smeared anti-commutators expectation values (as well as negativity) remain well-defined. Importantly, it is well-known that the behaviour of correlations of a massless field in 1+3 dimensions can be approximated by the momentum correlations\footnote{Considering the momentum as the fundamental field observable for a 1+1 massless scalar theory is equivalent to removing its zero-mode subspace.} of a 1+1 massless field~\cite{BenitoJormaUnruhState,derivativeJorma,Cong2019,ericksonBH}. Extrapolating this analogy, one expects that the exact MEM smearing profile $F^{\bd}$ in $1+3$ dimensions is not square integrable. Similarly, the analogy between 1+1 and 1+3 dimensions would imply that the exact MEM momentum profile $G^{\bd}$ is and $L^2$ function in 1+3 dimensions. These conclusions are also supported by the asymptotic behaviour of the expansion coefficients of the approximate MEM profiles discussed in the previous section. Moreover, our method yields regular profiles within a physically accessible subspace that captures the maximal two-mode distillable entanglement (as measured by the negativity) between two regions.

\subsection{Convergence of the Logarithmic Negativity}
Finally, we examine the convergence of the Logarithmic Negativity (LN) between the MEMs in regions $\Sigma_\tc{a}$ and $\Sigma_\tc{b}$ as the dimension of the associated symplectic subspace is increased.  The MEMs enter the calculation of the LN only through the covariance matrix of the corresponding smeared canonical observables. Hence, the convergence of these smeared correlation functions $\langle\{\hat{\Phi}(F_N^{\bd}),\hat{\Phi}(F_N^{\bd})\}\rangle$ and $\langle\{\hat{\Pi}(G_N^{\bd}),\hat{\Pi}(G_N^{\bd})\}\rangle$ in the vacuum state---rather than convergence of the smearing profiles $F_N^{\bd}$ and $G_N^{\bd}$ in the $L^2$ norm---is what ultimately determines the convergence of the LN.



Figure~\ref{fig:LN_vs_nmodes} shows the LN between the MEMs in the spheres $\Sigma_\tc{a}$ and $\Sigma_\tc{b}$ as a function of the number of modes $NL_\text{max}$ included in the truncated symplectic subspace, when the spheres are separated by different distances $d$. For all separations considered, the LN increases rapidly when only a small number of radial basis functions are included. As $N$ increases, the growth rate decreases substantially. The absence of a visible plateau at the largest $N$ for small separations (such as $d = 0.1R$) does not indicate that the LN diverges as $N \to \infty$. Instead, it reflects the fact that increasingly fine boundary structure needs to be resolved for shorter separations between the spheres. 
\begin{figure}[h]
    \centering
\includegraphics[width=0.75\linewidth]{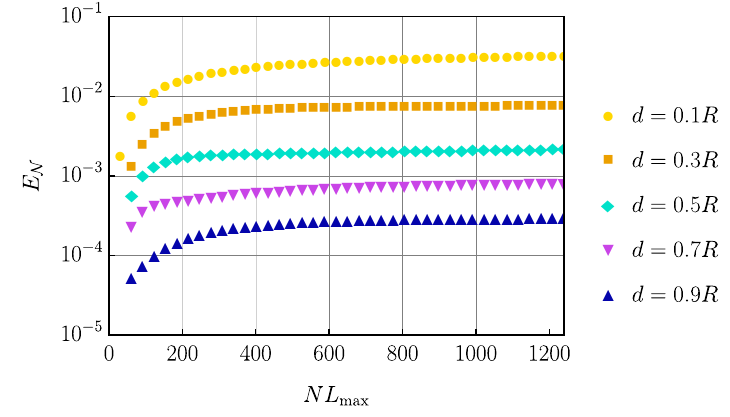}
    \caption{LN between the MEMs as a function of the number $N$ of basis functions included in the optimization in each region, for different separations between the spherical regions. Each point corresponds to fixing the truncation in the number of  spherical harmonics $l = 0, \dots, L_{\mathrm{max}}$ with $L_{\mathrm{max}} = 30$,  and progressively increasing the number of radial basis functions included in the expansion.  }
    \label{fig:LN_vs_nmodes}
\end{figure}

This expectation is further supported by the results in~\cite{Jason}, where the MEMs for a $1+1$ dimensional massless scalar field are obtained in the continuum limit in closed form. Although the corresponding mode profiles are not square-integrable (see previous subsection), the LN between the continuum MEMs is finite. Moreover, preliminary calculations using a similar truncation strategy~\cite{marcos_prep} reproduce the results in~\cite{Jason} for the LN between the MEMs once sufficiently many basis functions are included. 

In conclusion, the observed lack of $L^2$ convergence of $F^{\bd}$ merely reflects the regularity properties of the exact MEM field profile. The relevant convergence is instead better phrased in the context of homogeneous Sobolev spaces, in which the position and momentum MEMs naturally belong to $\dot H_{-1/2}(\mathbb{R}^3)$ and $\dot H_{1/2} (\mathbb{R}^3)$, respectively. The numerical MEMs seem to converge in the Sobolev norms (as evidenced by the fact that the smeared field and momentum anti-commutators remain finite), and the convergence of the LN can be understood as the physically relevant consequence of this fact.  

\end{document}